\documentclass{emulateapj}
\usepackage{graphicx,color}
\usepackage{amssymb,amsmath}
\usepackage[caption=false]{subfig}
\usepackage{ulem}
\usepackage[colorlinks,hyperfootnotes=false,citecolor=blue,linkcolor=blue]{hyperref}

\begin{document}
\shorttitle{SL models for four RELICS clusters}
\shortauthors{Cibirka et al.}

\slugcomment{Submitted to the Astrophysical Journal}

\title{RELICS: Strong Lensing analysis of the galaxy clusters Abell S295, Abell 697, MACS J0025.4-1222 and MACS J0159.8-0849}

\author{Nath\'alia Cibirka\altaffilmark{1}*, Ana Acebron\altaffilmark{1}, Adi Zitrin\altaffilmark{1}, Dan Coe\altaffilmark{2}, Irene Agulli\altaffilmark{1}, Felipe Andrade-Santos\altaffilmark{3}, Maru\v{s}a Brada\v{c}\altaffilmark{4}, Brenda Frye\altaffilmark{5}, Rachael C. Livermore\altaffilmark{6,8}, Guillaume Mahler\altaffilmark{7}, Brett Salmon\altaffilmark{2}, Keren Sharon\altaffilmark{7}, Michele Trenti\altaffilmark{6,8}, Keiichi Umetsu\altaffilmark{9}, Roberto Avila\altaffilmark{2}, Larry Bradley\altaffilmark{2}, Daniela Carrasco\altaffilmark{6}, Catherine Cerny\altaffilmark{7}, Nicole G. Czakon\altaffilmark{9}, William A. Dawson\altaffilmark{10}, Austin T. Hoag\altaffilmark{4}, Kuang-Han Huang\altaffilmark{4}, Traci L. Johnson\altaffilmark{7}, Christine Jones\altaffilmark{3}, Shotaro Kikuchihara\altaffilmark{11}, Daniel Lam\altaffilmark{12}, Lorenzo Lovisari\altaffilmark{3},  Ramesh Mainali\altaffilmark{5}, Pascal A. Oesch\altaffilmark{13}, Sara Ogaz\altaffilmark{2}, Masami Ouchi\altaffilmark{11,14}, Matthew Past\altaffilmark{7}, Rachel Paterno-Mahler\altaffilmark{7}, Avery Peterson\altaffilmark{7}, 
Russell E. Ryan\altaffilmark{2}, Irene Sendra-Server\altaffilmark{15,16},Daniel P. Stark\altaffilmark{5}, Victoria Strait\altaffilmark{4}, 
Sune Toft\altaffilmark{17} and Benedetta Vulcani\altaffilmark{6}}

\altaffiltext{1}{Physics Department, Ben-Gurion University of the Negev, P.O. Box 653, Be'er-Sheva 8410501, Israel\\ * nathcibirka@gmail.com}
\altaffiltext{2}{Space Telescope Science Institute, 3700 San Martin Drive, Baltimore, MD 21218, USA}
\altaffiltext{3}{Harvard-Smithsonian Center for Astrophysics, 60 Garden Street, Cambridge, MA 02138, USA}
\altaffiltext{4}{Department of Physics, University of California, Davis, CA 95616, USA}
\altaffiltext{5}{Department of Astronomy, Steward Observatory, University of Arizona, 933 North Cherry Avenue, Rm N204, Tucson, AZ, 85721, USA}
\altaffiltext{6}{School of Physics, University of Melbourne, VIC 3010, Australia}
\altaffiltext{7}{Department of Astronomy, University of Michigan, 1085 South University Ave, Ann Arbor, MI 48109, USA}
\altaffiltext{8}{Australian Research Council, Centre of Excellence for All Sky Astrophysics in 3 Dimensions (ASTRO 3D), Melbourne, VIC, Australia}
\altaffiltext{9}{Institute of Astronomy and Astrophysics, Academia Sinica, PO Box 23-141, Taipei 10617,Taiwan}
\altaffiltext{10}{Lawrence Livermore National Laboratory, P.O. Box 808 L-210, Livermore, CA, 94551, USA}
\altaffiltext{11}{Institute for Cosmic Ray Research, The University of Tokyo,5-1-5 Kashiwanoha, Kashiwa, Chiba 277-8582, Japan}
\altaffiltext{12}{Leiden Observatory, Leiden University, NL-2300 RA Leiden, The Netherlands}
\altaffiltext{13}{Geneva Observatory, University of Geneva, Ch. des Maillettes 51, 1290 Versoix, Switzerland}
\altaffiltext{14}{Kavli Institute for the Physics and Mathematics of the Universe (Kavli IPMU, WPI), The University of Tokyo, Chiba 277-8582, Japan}
\altaffiltext{15}{American School of Warsaw, Warszawska 202, 05-520 Bielawa, Poland}
\altaffiltext{16}{Department of Theoretical Physics, University of Basque
Country UPV/EHU, E-48080 Bilbao, Spain}
\altaffiltext{17}{Cosmic Dawn Center, Niels Bohr Institute, University of Copenhagen, Juliane Maries Vej 30, København ø, DK-2100, Denmark}

%

\begin{abstract}
We present a strong-lensing analysis of four massive galaxy clusters imaged with the Hubble Space Telescope in the Reionization Lensing Cluster Survey. We use a Light-Traces-Mass technique to uncover sets of multiply images and constrain the mass distribution of the clusters. These mass models are the first published for Abell S295 and MACS J0159.8-0849, and are improvements over previous models for Abell 697 and MACS J0025.4-1222. Our analysis for MACS J0025.4-1222 and Abell S295 shows a bimodal mass distribution supporting the merger scenarios proposed for these clusters. The updated model for MACS J0025.4-1222 suggests a substantially smaller critical area than previously estimated. For MACS J0159.8-0849 and Abell 697 we find a single peak and relatively regular morphology, revealing fairly relaxed clusters. Despite being less prominent lenses, three of these clusters seem to have lensing strengths, i.e. cumulative area above certain magnification, similar to the Hubble Frontier Fields clusters (e.g., A($\mu>5$) $\sim 1-3$ arcmin$^2$, A($\mu>10$) $\sim 0.5-1.5$ arcmin$^2$), which in part can be attributed to their merging configurations. We make our lens models publicly available through the Mikulski Archive for Space Telescopes. Finally, using Gemini-N/GMOS spectroscopic observations we detect a single emission line from a high-redshift $J_{125}\simeq25.7$ galaxy candidate lensed by Abell 697. While we cannot rule out a lower-redshift solution, we interpret the line as Ly$\alpha$ at $z=5.800\pm 0.001$, in agreement with its photometric redshift and dropout nature. Within this scenario we measure a Ly$\alpha$ rest-frame equivalent width of $52\pm22$ \AA\, and an observed Gaussian width of $117\pm 15$ km/s.

\vspace{0.05cm}
\end{abstract}


\section{Introduction}
\label{intro}
Our current understanding of early cosmic history is based on two main pictures: the first provides a view of the early Universe through measurements of the Cosmic Microwave Background (CMB) \citep{Komatsu2011,Planck2016b}, last scattered about 400,000 years after the Big Bang, while the second relies on observations of the first galaxies, which formed a few hundred million years later \citep{Rees1998,Barkana2001}. At that era the Universe was filled with neutral hydrogen, which was then gradually reionized by $z\sim6$ \citep{Fan2006,Robertson2015,Dijkstra2014}. During this reionization epoch, lasting only a few hundred Myr, the Universe went through a particularly rapid evolution \citep[see][for reviews]{Stark2016,Zaroubi2013}.

\begin{table*}
	\caption{Properties of RELICS clusters considered in this work}            
	\label{table:1}      
	\centering   
    {\renewcommand{\arraystretch}{1.2}
	\begin{tabular}{c c c c c}        
		\hline\hline                 
		Cluster & R.A. & Dec& Redshift & Planck SZ mass \\  
		&[J2000]&[J2000]&&M$_{500}$ [$10^{14}M_\odot$]   \\
		\hline          
        MACS J0025.4-1222 & 00:25:29 & -12:22:54 & 0.586 & - \\ 
        MACS J0159.8-0849 & 01:59:54 & -08:51:32 & 0.405 & 7.20\\
        Abell S295 & 02:45:28 & -53:02:32 & 0.300 & 6.78\\ 
        Abell 697 & 08:42:59 & +36:21:09 & 0.282 & 11.00\\     
		\hline\hline                                
	\end{tabular}}
    \vspace{0.3cm}
\end{table*}

One way to explore the reionization epoch is based on the statistics of high-redshift galaxies, expressed in terms of the galaxy luminosity function (\citealt{Bromm2011} and references therein, \citealt{McLure2013,Finkelstein2015,Bouwens2017,Livermore2017}), which can be then translated to an ionizing photon budget \citep[][]{Atek2015,Mason2017,Morales2010,Robertson2015}. This is of particular interest, since it is currently uncertain whether galaxies could fully account for reionization. The
detection of increasing numbers of high-redshift galaxies
is thus crucial. Observing high-redshift galaxies, however, is
challenging. Moreover, current observations of galaxies at $z \gtrsim 6$ correspond typically to the brighter ($L \gtrsim L^{\ast}$) objects, hence conclusions about the faint end of the luminosity function, representing the more abundant population of galaxies, cannot be directly obtained. Over the past decade there have been growing efforts to observe samples statistically representative of the underlying, fainter high-redshift galaxy population that may have been responsible for reionization. Significant progress has been achieved with deep and high resolution observations carried out by the \textit{Hubble Space Telescope} (HST) under various programs (e.g. \citealt{Beckwith2006,Scoville2007,Grogin2011,Bradley2012,Ellis2013,Bouwens2015,Finkelstein2016,Livermore2017}; see \citealt{Stark2016} for a review). 

A second route to studying galaxies at high redshifts and their contribution to reionization is via spectroscopy. However, spectroscopy of Lyman-$\alpha$ or other UV metal lines from very distant objects in the reionizaton era is extremely challenging \citep{Stark2010,Pentericci2011, Schenker2014,Schmidt2016,Laporte2017,Hoag2017}. As bright galaxies at high redshifts are scarce, there has been a growing need to discover more high-z candidates that are apparently bright enough to be studied spectroscopically \citep[e.g.][]{Roberts-Borsani2016}.

The discovery of galaxies at high redshift is enhanced by strong gravitational lensing. Acting as natural telescopes, massive galaxy clusters magnify background sources that are intrinsically faint and would otherwise remain undetectable \citep[e.g.,][]{Zheng2012,Coe2013,Zitrin2014}. Indeed, recent cluster lensing campaigns have been detecting increasing numbers of magnified high-redshift sources \citep[e.g.,][]{Bradley2014, Monna2014, Jauzac2015, Huang2016, Zitrin2017}, enabling the community to probe the fainter-end of the luminosity function, up to $z \sim 9-10$ \citep{McLure2013,Oesch2014,Atek2015, McLeod2016, Livermore2017, Bouwens2017, Ishigaki2017, Kawamata2017}. 

Apart from the magnification effect useful for studying lensed galaxies, gravitational lensing also provides a unique way to map the total mass content of clusters, including both the baryonic and dark matter (DM) components \citep{Schneider1992, Clowe2006, Bartelmann2010, Kneib2011}.

Following the success of previous lensing surveys like the Cluster Lensing and Supernovae Survey with Hubble \citep[CLASH][]{Postman2012, Coe2013, Bradley2014} and the Hubble Frontier Fields \citep[HFF][]{Coe2015, Lotz2017}, the Reionization Lensing Cluster Survey (RELICS; Coe et al. in prep) was designed to study 41 clusters and reveal high-redshift galaxies ($z \sim 6-12$), in particular, apparently bright, highly magnified examples that could be followed up spectroscopically from the ground or with the James Webb Space Telescope (JWST) \citep[e.g.][]{Salmon2017}.

In order to account for the strong lensing (SL) effect and correctly interpret the results, such as the intrinsic properties of lensed and high-redshift galaxies, it is necessary to derive a detailed lens model, which is our goal here. We present a SL analysis of four clusters observed in the framework of the RELICS\footnote{\url{relics.stsci.edu}} program. We briefly introduce the program and the observations in Section \ref{sec:data}, where we also provide an overview of the clusters analyzed in this work. In Section \ref{sec:lens_model} we describe the adopted strong-lens modeling technique. We present the individual lens modeling for each of the clusters in Section \ref{analy}. Our findings are presented and discussed in Section \ref{sec:discuss}, and summarized in Section \ref{sec:summ}. 

Throughout this work we adopt the standard ${\Lambda\mathrm{CDM}}$ flat cosmological model with a Hubble constant of $H_0 = 70\ \rm{km s}^{-1}\ \rm{Mpc}^{-1}$ and ${\Omega_{\rm{M}}}=0.3$. Magnitudes are quoted in the AB system. Errors correspond to the $68.3\%$ confidence level unless otherwise specified. The errors we quote for the Einstein radius and mass throughout are 10\% and 15\%, respectively. These were found to also encompass typical differences between different lens modeling techniques \citep[e.g.][]{Zitrin2015}.


\section{Data and observations} 
\label{sec:data}

Our strong lensing analysis is based on observations from the RELICS program (PI: D. Coe). The RELICS is a 188-orbit Hubble Space Telescope (HST) Treasury Program (GO 14096; PI: Coe), that has observed 41 galaxy clusters. The goal of the project is to analyze these massive clusters and find magnified, high-redshift galaxies. A Spitzer imaging campaign (PI: Bradaˇc, PI: Soifer) of more than 500 hours accompanies the program, and is intended to improve the search for high-redshift galaxies and complement the studies on the observed galaxy properties. A detailed description of the RELICS program and the sample selection will be presented in an upcoming paper (Coe et al., in preparation).

The selection of clusters for the program was mainly based on the Sunyaev Zel’dovich effect (SZ). The first half of the sample consists of 21 of the 34 most massive Planck clusters \citep{Planck2016} for which the SZ-masses are similar to or greater than those of the HFF clusters and for which no HST/IR imaging was available. The remaining clusters were selected based on several criteria such as mass estimates from X-ray \citep[MCXC][]{piffaretti2011,mantz2010} and weak lensing \citep{sereno2014,applegate2014,vonderlinden2014,umetsu2014,hoekstra2015}, SZ mass estimates from South Pole Telescope \citep[SPT,][]{Bleem2014} and Atacama Cosmology Telescope \citep[ACT,][]{Hasselfield2013} data, as well as following an analysis of clusters from the Sloan Digital Sky Survey \citep[SDSS,][]{wong2013,wen2012}. This combined selection based primarily on mass was aimed at increasing the probability of detecting high-redshift galaxies since, overall, massive clusters tend to be more efficient lenses. 

The 41 clusters were observed in the optical and near-infrared. The images were obtained using three Advanced Camera for Surveys filters (ACS -- F435W, F606W, F814W) for one orbit each, and four Wide Field Camera 3 filters (WFC3/IR -- F105W, F125W, F140W, F160W), for half an orbit each. In total each cluster was thus observed for 5 orbits, with the exception of some cases where HST archival data was available. All sub-exposures in each filter were combined to create a deep image in that band. The final drizzled images were produced in pixel scales of $0.03"$ and $0.06"$, after matching the filters to the same pixel frame and correcting the astrometry to the Wide-field Infrared Survey Explorer (WISE) point source catalog \citep{Wright2010}. An interval of about one to two months between observations of the same cluster was typically designed to allow the identification of variable sources.

\begin{figure*}
	\centering
	\includegraphics[width=0.75\linewidth,trim=0cm 0cm 0cm 0cm,clip]{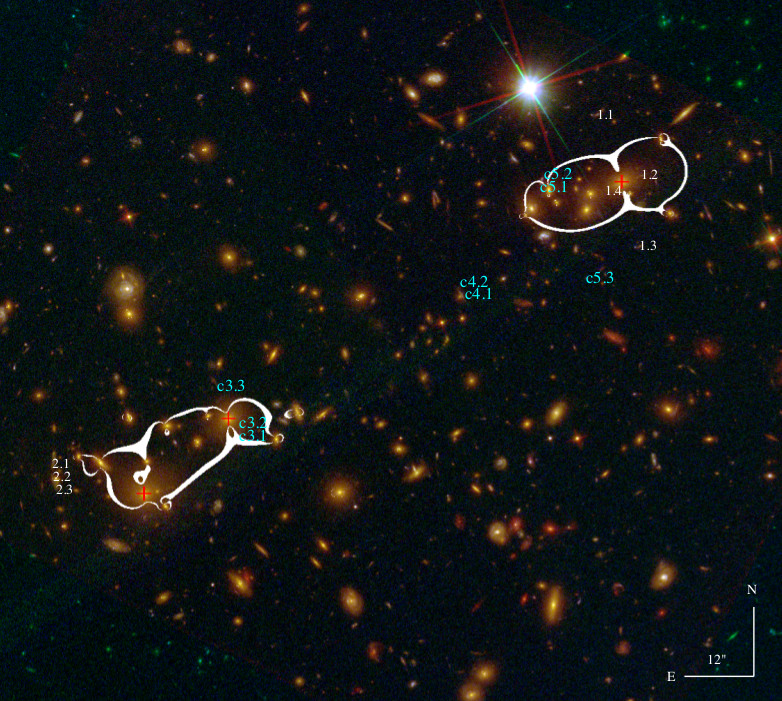} 
 	\caption{RELICS color image of MACS J0025.4-1222 (Blue=F435W, Green=F555W+F814W, Red=F105W+F125W+F140W+F160W). The white contours show the critical curves from our best-fit model, for a source at $z_{\rm{spec}} = 2.38$ (system 1). Multiple images considered in the modeling are labeled in white, while candidate systems (not used in the fit) are shown in cyan (see also Table \ref{table:0025}). The positions of the three main BCGs are marked with a red cross. We sometimes refer to the colors of multiple images with respect to these composite images.}
	\label{macs0025cc}
\end{figure*} 

\begin{figure*}
	\centering
	\includegraphics[width=0.75\linewidth]{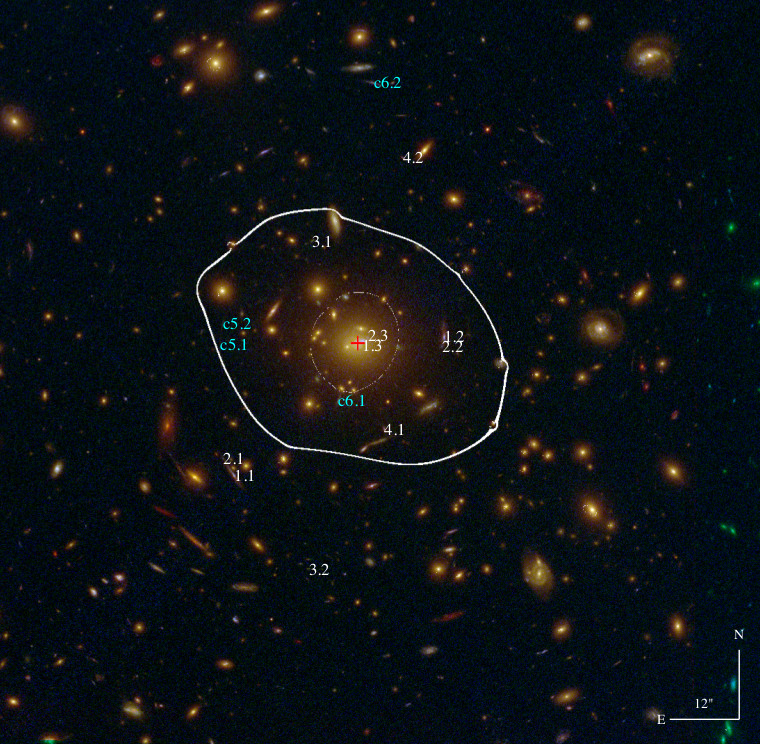} 
 	\caption{RELICS color image of MACS J0159.8-0849 (Blue=F435W, Green=F606W+F814W, Red=F105W+F125W+F140W+F160W). The white contours show the critical curves from our best-fit model, for a source at $z = 1.55$. Multiple images considered in the modeling are labeled in white, while candidate systems (not used in the fit) are shown in cyan (see also Table \ref{table:0159}). The BCG position is marked with a red cross.}
	\label{macs0159cc}
\end{figure*} 

\begin{figure*}
	\centering
	\includegraphics[width=0.75\linewidth,trim=0cm 0cm 0cm 0cm,clip]{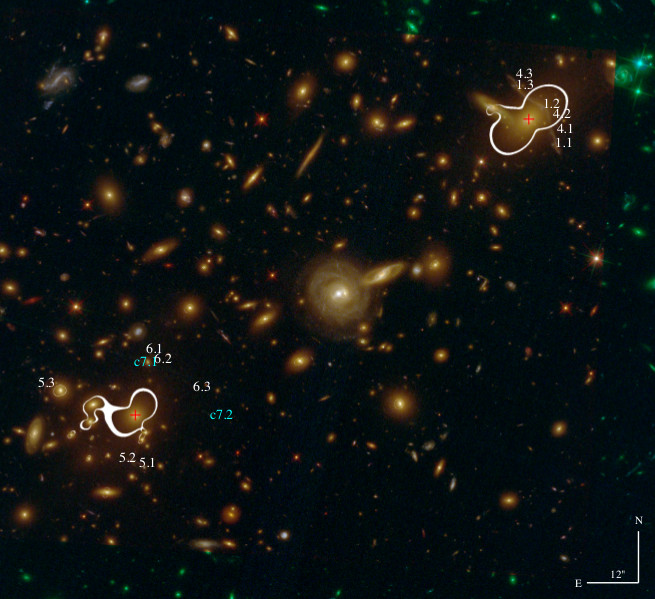} 
 	\caption{RELICS color image of Abell S295 (Blue=F435W, Green=F606W+F814W, Red=F105W+F125W+F140W+F160W). The critical curves for our best-fit model are shown in white, corresponding to a source at $z = 1$. Multiple images used in the modeling are labeled in white, while candidate systems (not used in the fit) are shown in cyan (see also Table \ref{table:S295}). For clarity we have omitted subset-systems 2 and 3 which are located in the giant arc very close to systems 1 and 4. The positions of the two BCGs are marked with red crosses.}
	\label{abellS295cc}
\end{figure*}

\begin{figure*}
	\centering
	\includegraphics[width=0.75\linewidth,trim=0cm 0cm 0cm 0cm,clip]{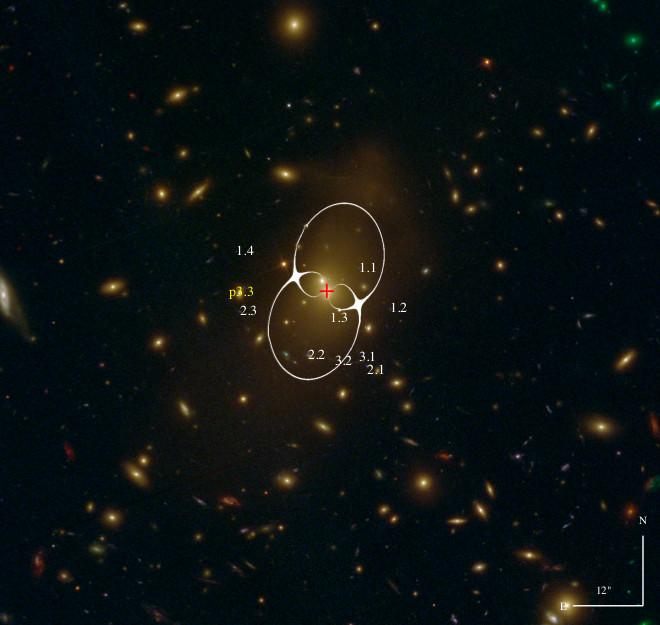} 
 	\caption{RELICS color image of Abell 697 (Blue=F435W, Green=F606W+F814W, Red=F105W+F125W+F140W+F160W). The white contours show the critical curves for our best-fit model, for a source at $z = 2$. Multiple images displayed in white are labeled according to Table \ref{table:697}. The position of a third counter image for system 3 predicted by the best-fit model is indicated in yellow. The BCG is marked with a red cross.}
	\label{abell697cc}
\end{figure*}

RELICS has delivered reduced HST images, photometry, and photometric redshift catalogs for all of the observed fields \citep[see also][]{Cerny2017}. Source catalogs are produced by running \textsc{SExtractor} \citep{Bertin1996} in dual-image mode, where a weighted, stacked dataframe combined from all near-infrared, and one combined from all optical and near-infrared bands, are used as reference. Most of the multiple images we identify or consider are detected by \textsc{SExtractor} with the initial chosen parameters, but for some fainter or blended objects we independently rerun SExtractor after manually varying the parameters until our objects of interest are detected and measured (typically this entailed increasing the number of deblending sub-thresholds by a factor $\times 2 - \times4$, lowering the deblending minimum contrast by a factor $\times 5 -\times10$, or changing the background mesh size by a factor $\times 2$). Photometric redshift estimates (also referred to here as photo-z) are then derived using the \textit{Bayesian Photometric Redshifts} algorithm \citep[BPZ,][]{Benitez2000, Benitez2004, Coe2006} with 11 templates for the spectral energy distribution including ellipticals, late types and starbursts \citep{Benitez2014, Rafelski2015}. RELICS also generates several color-composite images for each cluster, which we use here, constructed from optical and near-infrared bands as indicated in Figures \ref{macs0025cc}-\ref{abell697cc} (we sometimes refer to the colors of multiple images with respect to these composite images). The reduced data and catalogs are available for the community through the Mikulski Archive for Space Telescopes (MAST)\footnote{\url{https://archive.stsci.edu/prepds/relics/}\label{mast}}.

Out of the four clusters analyzed in this work (Table \ref{table:1}), two are Abell clusters \citep{abell1989}, one of which belongs to the supplementary Abell catalog. The other two clusters are part of the MAssive Cluster Survey \citep[MACS,][]{Ebeling2001}. Throughout the work we discuss the four clusters in an increasing Right Ascension order. 

The first cluster we analyze is MACS J0025.4-1222 (the ``baby bullet'', MACS0025 hereafter). This is a massive, major merger cluster at $z = 0.586$ with the collision taking place approximately in the plane of the sky \citep{bradac08,ma10}. It is also among the most X-ray luminous clusters at $z > 0.5$ \citep{ebeling07}. A clear separation between the DM component and the intracluster gas can be observed, in a similar way to the ``Bullet Cluster" \citep[1E 0657–56,][]{clowe04}. While both the galaxy and DM components show a bimodal distribution corresponding to two subclusters with similar masses, the gas shows a single peak located between the two substructures. \cite{bradac08} studied the distribution of the different components using strong and weak lensing information obtained from multi-color HST images (observed with the ACS - F555W and F814W filters and WFPC2 - F450W filter) and X-ray data from Chandra. They found that the observed offset between the hot gas and the mass distribution based on the galaxies and lensing maps are in agreement with what is expected for collisionless cold DM (CDM). Approximately 200 objects in the cluster were targeted as part of the DEIMOS/KECK spectroscopic campaign for the MACS survey \citep{ebeling07}, providing a precise measurement of the cluster redshift and its velocity dispersion. Additionally, multiple-image candidates selected from the HST data available at the time (GO 9722, 10703, PI: Ebeling; GO 11100, PI: Brada{\v c}) were observed with LRIS/KECK \citep{bradac08}, and a couple of lens models for this cluster have been published prior to RELICS data \citep{bradac08,Zitrin2011}.

The second cluster is MACS J0159.8-0849 (MACS0159), a hot and luminous X-ray cluster at $z=0.406$, showing a regular X-ray morphology \citep{maughan2008}. This cluster is part of the sample comprising the 34 brightest MACS clusters \citep{ebeling10}. MACS0159 is also one of only two RELICS clusters that are part of the \cite{wong2013} rank of prominent lenses, based on SDSS luminous red galaxy data. The presence of an extended X-ray source detected as a filamentary structure was reported by \cite{kotov2006}, who analyzed the mass-temperature relation of clusters at $z \sim 0.5$. An extended and diffuse emission around the BCG was also observed in the radio by \cite{Giacintucci2014}, in an analysis searching for radio minihalos in a sample of X-ray luminous clusters. For this cluster we made use of the available archival HST data from GO programs 11103 and 12166 (PI: Ebeling).

Abell S295, the third cluster we analyze, \citep[AS295,][]{abell1989}, is located at $z = 0.30$. This cluster is also known as SPT-CL J0245-5302 and ACT-CL J0245-5302, and is part of the sample comprising the 26 most massive clusters detected by their SZ effect in the SPT Survey \citep{williamson2011}. \cite{edge1994} reported the discovery of a giant arc, dubbed ``the Mexican hat", in the North-West region of the cluster, detected during optical follow-up of the ROSAT All-Sky Survey as part of an ESO Key Program. At that time the data revealed the presence of three knots in the arc. AS295 was also studied by \cite{Menanteau2010}, who discussed optical and X-ray properties of the first ACT cluster sample and noted that this cluster is likely a merger. Further works which have included AS295 are the ones by \cite{Ruel2014} and \cite{Bayliss2016}, who presented and analyzed spectroscopic data for a sample of clusters detected in the SPT survey. For this cluster we used the available archival HST data from GO program 13514 (PI: Pacaud).     

The last cluster we analyze here is Abell 697 \citep[A697 hereafter,][]{abell1989, Crawford1995}, a rich and massive cluster located at $z=0.282$. This cluster is the tenth-highest Sunyaev-Zel'dovich mass cluster in the Planck catalog and is part of the ROSAT Brightest Cluster Sample \citep[BCS,][]{ebeling1998}, being a hot and luminous cluster in the X-ray. \cite{kempner2001} suggested the presence of a diffuse cluster-scale radio emission, later classified as a giant radio halo (\citealt{Venturi2008}). Subsequent radio, optical and X-ray analyses hint that A697 is the result of a complex multiple merger history occurring along the line-of-sight \citep{girardi2006,macario2010}. This finding is supported by the first lens model of this cluster based on data from the Keck Observatory \citep{metzger2000}. The authors observed a gravitationally lensed arc to the south of the big cD central structure, which entailed a very elliptical potential in the derived SL model. 

\subsection{Spectroscopic Observations}\label{gmos}

We obtained Gemini-N GMOS spectroscopic observations for A697 (program ID: GN-2018A-Q-903, PI: Zitrin), primarily targeting multiply imaged galaxies listed in Figure \ref{abell697cc} (see also Table \ref{table:697}). The mask included slits placed on images 1.1, 1.4, 2.3 and 3.1, so that at least one image from each multiple-image system is observed. Observations were carried out in queue mode on the night of 2018-03-18. Raw weather conditions for execution were 50-percentile cloud cover, 20-percentile background brightness, and 70-percentile image quality. From acquisition stars on our mask we measure an average seeing of $0.55\arcsec$, and online webcam data\footnote{http://mkwc.ifa.hawaii.edu/current/cams} suggest some possible cloud coverage. A total integration time of 3872 sec was achieved using four 968 sec long exposures, in a Nod \& Shuffle mode with a nod of $\pm0.75$ arcseconds. We adopted the GG455 filter and R400 grating, and a central wavelength (CW) of 700 and 710 \AA\ (two CWs were used to fill gaps between the CCDs). The resulting coverage was $\sim 4700-9500$ \AA\, depending on the position of the slit in the mask, with a resolution of 1.516 \AA/pixel. No pre-imaging was required given the available HST images. Data were retrieved through the Gemini Observatory Archive and processed with the Gemini IRAF package using the standard procedure that includes bias, dark, flats, and wavelength calibration. The processed images were combine to a final two-dimensional spectra for each slit, which was then also extracted to one-dimension. In each slit of the four raw images we also planted artificial sources to track the reduction progress, given the lack of a continuum emission, to verify that the offsets assigned in various reduction steps match the expectations. 

In addition to multiple images, we also placed slits on high-redshift candidates from \citet{Salmon2017} (see Table \ref{table:highzcan} here), although we note that the relatively short exposure time was dictated by multiple images, typically brighter than high-redshift candidates which were secondary targets here.

We did not identify any prominent emission line in the three multiple-image systems, and so, we do not obtain new spectroscopic redshifts for the modeling. However, we do detect a prominent line for one of the high-redshift galaxies, namely Abell697-0636 listed in Table \ref{table:highzcan}, which we interpret as Ly$\alpha$, yielding a spectroscopic redshift of $z_{Ly\alpha}=5.800 \pm 0.001$ (another possible interpretation is [O II] at a redshift of $\sim 1.2$). The final one and two-dimensional spectra are shown in figure \ref{spec7}, and presented in more detail in \S \ref{line}.

\section{Lens modeling formalism} 
\label{sec:lens_model}

We adopt a Light-Traces-Mass approach to model the projected central mass distribution of the clusters. One of the advantages of this approach is its predictive power, guiding the identification of multiple-image systems. The method relies on the simple assumption that the distribution of the observed galaxies traces the overall DM distribution of the cluster. In the following we give an overview of the method. For a detailed description we refer to \citet[][see also \citealt{broadhurst2005}]{zitrin2009,Zitrin2015}.

The modeling starts with the construction of a catalog of cluster members using the red sequence method \citep{Gladders2000}. For each cluster, we use the magnitudes measured from the F606W and F814W filters to draw the color-magnitude diagram, and identify the location of the red sequence. We typically consider galaxies to be cluster members when lying within $\pm0.3$ mag of the sequence, selecting galaxies down to 23 AB. We follow this first selection with a visual inspection of the HST color image and manually exclude objects with a doubtful morphology or color. Similarly, we sometimes manually include galaxies that may have been lost in the initial selection. Additionally, the flux of a few galaxies, typically, has to be reduced manually; especially galaxies that are designated as cluster members but show bright spiral structure, and so have in practice a significantly lower M/L than the one effectively adopted for cluster ellipticals. We discuss this point in some more detail in \S \ref{sec:discuss}.

The next step is to construct the mass distribution based on the final catalog of selected cluster galaxies. We start by assigning a power-law surface mass-density radial profile to each galaxy:

\begin{equation}
\Sigma(r) = K r^{-q}.
\label{powerLaw}
\end{equation}

The amplitude of this profile, $K$, is linearly scaled with the observed luminosity. The power-law exponent $q$, is the same for all galaxies, and is a free parameter of the model. The enclosed mass of a galaxy within an angular distance $\theta$ is then given by:

\begin{equation}
M(<\theta) = \frac{2 \pi K}{2-q}(D_l \theta)^{2-q},
\end{equation}

and its deflection field is:

\begin{equation}
\alpha(\theta) = \frac{4GM(< \theta)}{c^2 \theta} \frac{D_{\rm{ls}}}{D_{\rm{l}} D_{\rm{s}}} ,
\end{equation}

\noindent where $D_{\rm{ls}}$, $D_{\rm{l}}$ and $D_{\rm{s}}$ are the angular diameter distances between lens-source, observer-lens and observer-source respectively. The last equation can be rephrased and written as:

\begin{equation}
\alpha(\theta) = K_q F \theta^{1-q},
\end{equation}

where $F$ is the measured flux, and $K_q$ is a constant that depends on the power-law index $q$ and encompasses all the previous constants and proportion relations. The deflection angle at a certain position due to all cluster member galaxies, $\vec{\alpha}_{\rm{gal}}$, is given by a linear sum of all individual galaxy contributions. The sum of all galaxy mass-density distributions therefore defines the galaxy component of the model, which should now be supplemented by a DM distribution. 

As the DM distribution should be smoother than the contribution of the galaxies, we apply a 2D Gaussian smoothing to the co-added galaxy component. We define $S$ to be the width of the 2D Gaussian and the second free parameter of the model. The deflection field resulting from the smooth component is defined as $\vec{\alpha}_{\rm{DM}}$. The overall normalization (essentially $K_{q}$) and the relative weight of the galaxies to the DM component (which we denote $K_{gal}$), are two other free parameters of the model.

Since some difference is expected between the distribution of galaxies and DM, a two-parameter external shear is added to the deflection field for further flexibility (also, the external shear effectively introduces ellipticity to the magnification map). The external shear is described by its amplitude and its position angle, which are also free parameters of the model. The deflection field from the external shear is marked as $\vec{\alpha}_{\rm{ex}}$. The basic model has thus a total of six global free parameters.

The total deflection field is then obtained by adding the three components up and accounting for their relative contributions:

\begin{equation}
\vec{\alpha}_T (\vec{\theta}) = K_{\rm{gal}}\vec{\alpha}_{\rm{gal}}(\vec{\theta}) + (1 - K_{\rm{gal}})\vec{\alpha}_{\rm{DM}}(\vec{\theta})+\vec{\alpha}_{\rm{ex}}(\vec{\theta}).
\end{equation}

The model can be further improved, typically, by allowing the weight of few central, brightest cluster galaxies (BCGs) to vary as free parameters. Similarly, a core radius and ellipticity can be introduced to these BCGs, in order to improve the fit. In addition, redshifts of systems lacking spectroscopic measurements can be left as free parameters and optimized in the fitting routine.

We use a Monte Carlo Markov Chain (MCMC) code with several thousand steps to obtain the best fit model, adopting a $\chi^2$ criteria:

\begin{equation}
\chi^2 = \sum\limits_{i=1}^{n} \dfrac{( x_i^{\rm{m}} - x_i^{\rm{obs}})^2 + ( y_i^{\rm{m}} - y_i^{\rm{obs}})^2}{\sigma_{i}^2} \mathrm{,}
\end{equation} 

\noindent where $x_i^{\rm{obs}}$, $y_i^{\rm{obs}}$ are the observed multiple-image positions, $x_i^{\rm{m}}$, $y_i^{\rm{m}}$ the corresponding coordinates predicted by the model and $\sigma_{i}$ the positional uncertainty. This minimization quantifies the distance of the multiple image positions inferred by the model with respect to the observed ones (we assume here a positional uncertainty of $0.5"$, e.g. \citealt{Newman2013}). 
To assess the uncertainty of the best-fit model we consider the \textit{rms} of the reproduced images in the image plane:

\begin{equation}
rms = \sqrt[]{\frac{1}{N} \sum_{i=1}^n (x_i^{\rm{m}} - x_i^{\rm{obs}})^2 +(y_i^{\rm{m}} - y_i^{\rm{obs}})^2} ,
\end{equation}

\noindent given the predicted ($x_i^{\rm{m}}, y_i^{\rm{m}}$) and observed ($x_i^{\rm{obs}},y_i^{\rm{obs}}$) positions and the total number $N$ of multiple images.

With only six global parameters, and relying on a simple assumption, the Light-Traces-Mass method is a powerful tool to identify multiple images and probe the cluster mass distribution. 

We start by constructing a preliminary model (using typical parameter values) based on the red-sequence cluster member selection and photometry. Thanks to the LTM assumption, this initial guess is sufficiently successful to guide the identification of multiple images, which are ultimately identified by eye, based on the morphology, position, and color information obtained from the HST images and compared with the model's prediction. We search for new multiple-image candidates predicted by the model in an iterative way, refining the model in each step. For the final model, we only use the most secure systems, i.e. those that agreed with the model's prediction and are conspicuously similar to the prediction and to each other in terms of symmetry, color, internal details, photometric redshift, etc. 

Most of our multiple-image systems do not have spectroscopic redshifts, and for these cases our modeling primarily uses the photometric redshift estimates from BPZ, averaged over all images of a given system. We checked individually the solutions given by BPZ and in cases where the photo-z of multiple images from the same system does not match, we manually assigned a higher weight to brighter and isolated images when determining the adopted system's photo-z. In addition, typically, we set for each cluster one multiple-image system with a very good photo-z estimate, and leave the redshift of the other systems to be optimized in the minimization around the best-fit value of their photo-z probability distribution function. This will be detailed for each cluster individually in \S \ref{analy}. 


\begin{figure*}
\centering          
\includegraphics[width=0.89\columnwidth]{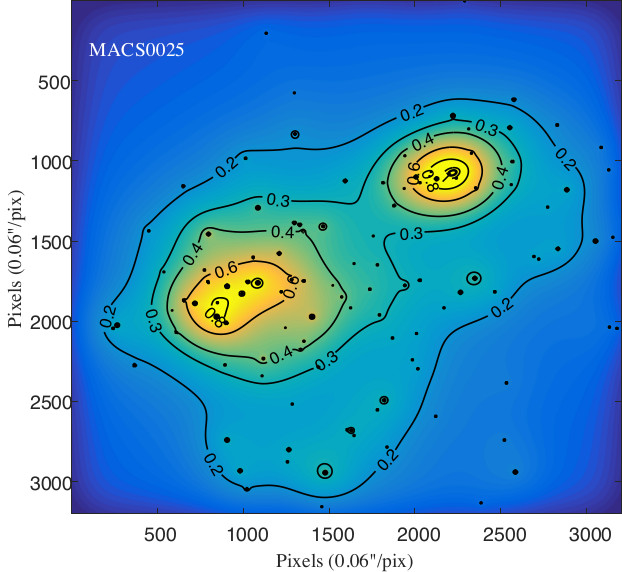}
\includegraphics[width=0.92\columnwidth]{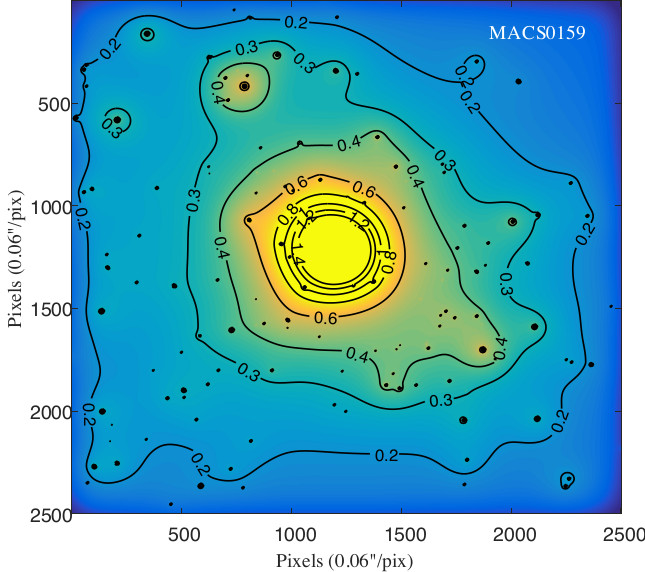}\\
\includegraphics[width=0.9\columnwidth]{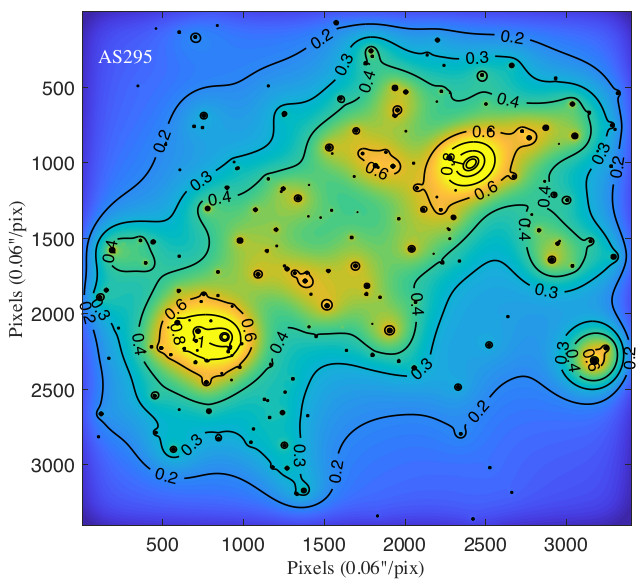}
\includegraphics[width=0.92\columnwidth]{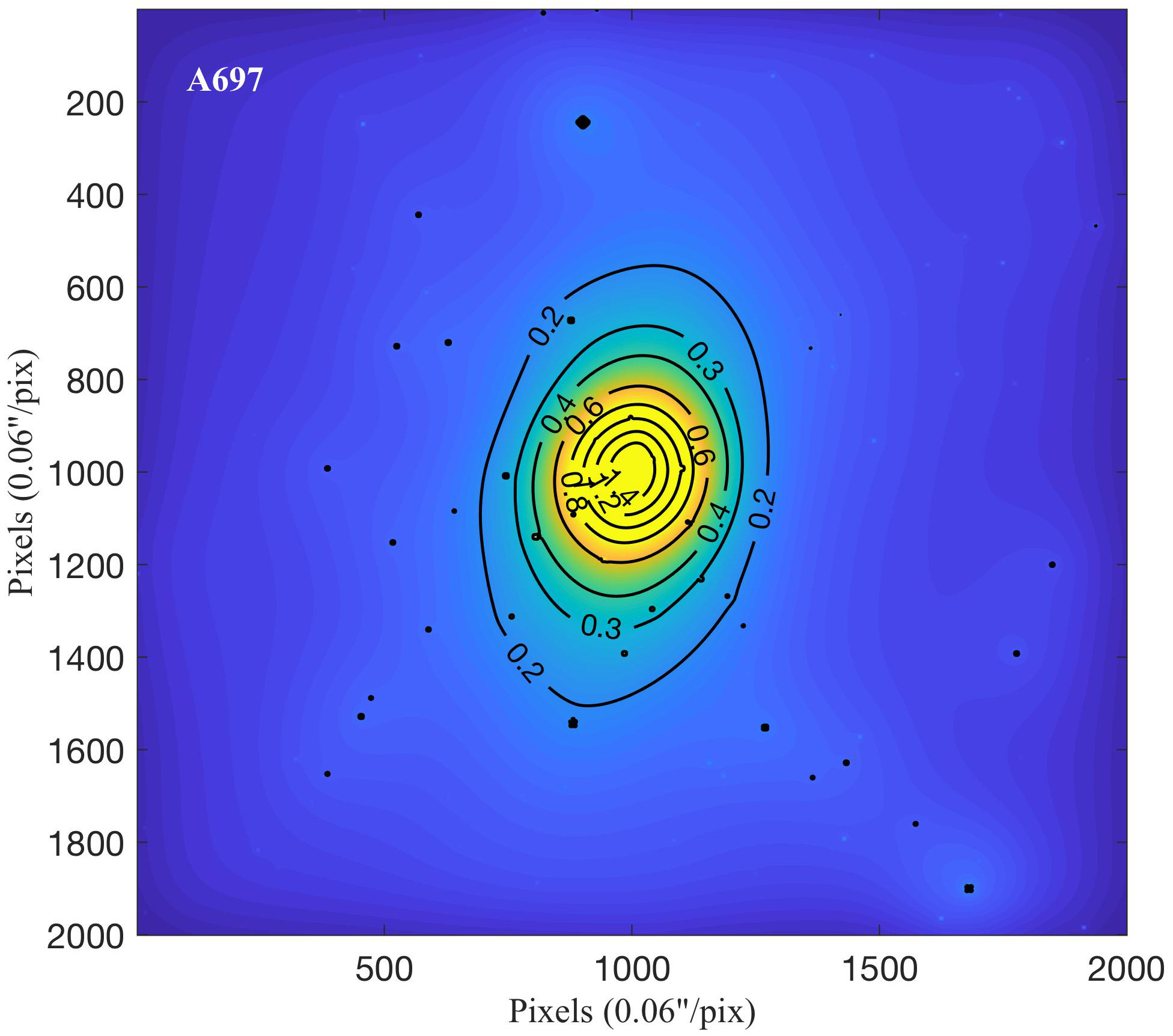}
\caption{Convergence map of each cluster produced from our best-fit models. The maps show the projected surface mass-density distribution in units of the critical density for lensing, for a source at $z = 2$. The orientations are the same as in Figures 1-4.} 
\label{kappas}
\end{figure*}

\section{Individual lens models}
\label{analy}

In the following subsections we present for each cluster the strong lens modeling and some of its immediate results. 

The projected mass-density distributions from the best-fit models, for a source located at $z_{s} = 2$, are shown in Fig. \ref{kappas}, and the corresponding mass-density profiles in Fig. \ref{profs}. We define the mass-density distribution of the systems in terms of the convergence, $\kappa$, a dimensionless surface density given by:

\begin{equation}
\kappa(\vec{r}) = \frac{\Sigma(\vec{r})}{\Sigma_{\rm{cr}}},
\end{equation}

\noindent where $\Sigma_{\rm{cr}}$ is the critical density for lensing given by:
\begin{equation}
\Sigma_{\rm{cr}} = \frac{c^2}{4 \pi G} \frac{D_s}{D_{ls}D_l}\ ,
\end{equation}
and $\Sigma(\vec{r})$ is the projected, surface mass density.

High-redshift magnification maps, scaled to a source redshift of $z_{s} = 9$, are shown in Fig. \ref{mags}. A list of the multiple image systems used as constraints can be found in Appendix \ref{appenTab}. We also list therein candidate multiple images (labeled with a \textit{c}). These are images or systems whose identification was more ambiguous and were not included as SL constraints (for example, images with similar color and/or morphology, but notable discrepancy between the observed and model-predicted location or orientation).

\subsection{MACS J0025.4-1222}


Our model for MACS0025 is constructed based on two systems of multiple images. System 1 forms at the North-Western subcluster and is composed of four counter images showing a two-component morphology, which includes a blue bright spot and a diffuse feature. Three images of this system were identified in \cite{bradac08} as system AB. \citet{bradac08} also measured a spectroscopic redshift for this system of $z = 2.38$ using Keck/LRIS data. A second lens model by \citet{Zitrin2011} has predicted a fourth image for this system, which they also identified in the data and is used here.

The second system we consider here was labeled system C in \cite{bradac08}, and we define three multiple images located on the east side of the main BCGs as constraints. This system was also spectroscopically targeted by \citet{bradac08}, but did not yield a spectroscopic redshift. \citet{bradac08} estimated a photometric redshift of $z_{phot} = 1.0^{+0.5}_{-0.2}$ for this system, due to the $4000$ \AA\ break possibly lying between the F555W and F814W bands. The BPZ estimate from RELICS gives a higher $z_{phot} = 3.8$ (corresponding to the value for the counter images denoted as 2.2 and 2.3 here), a redshift which we fix for this system in our modeling. We note that this estimate uses information from the seven HST band observations available for RELICS, while \cite{bradac08} used the information from the F450W/F555W/F814W bands available at the time. This choice of fixing both redshifts in our model was made in order to constrain both mass clumps. 

\citet{bradac08} also included a third system in their SL model, labeled system D. Here we do not include this as constraints. Based on the new information from our multi-band observations, the northern counter image in the region of the second BCG (labeled c3.3) does not seem to match the southern arc-shaped images (c3.1-2) in terms of color. In addition, our model does not precisely predict the northern counter image, nor the unexpected orientation of the southern arc. So we designate this system as a candidate system only (labeled as system c3). We note also that \citet{Zitrin2011} included a fourth two-image system that their older model agreed with. We decided to discard this system from our analysis as it appears to be a single elongated image with no counter image predicted by our model, and only mention for completeness as a candidate system (c4). 

In our modeling we include the weight of the north-western BCG as a free parameter to be optimized, allowing it to vary with respect to the original M/L ratio. We set the ellipticity and position angle for this BCG to the values given by \textsc{SExtractor}. The resulting critical curves, computed for a source at $z = 2.38$, and the location of the multiple images, can be seen in Fig. \ref{macs0025cc}. 

The multiple image reproduction is shown in Fig. \ref{stamp_0025}. Our model precisely predicts the four counter images of system 1 with respect to the position and orientation. The prediction for system 2 returns an arc-shaped image with the orientation matching the observed disposition of the system counter images, although the specific position of the predicted images is less accurate, and likely influenced by the local cluster member around which the arc partially revolves. We acknowledge that given the lack of internal details in system 2, its exact configuration remains ambiguous. The final \textit{rms} of the best-fit model is $0.57"$.

Due to the small number of constraints, i.e., multiple images that could be reliably identified, the model for MACS0025 is in a sense, a simple model, with a minimum number of free parameters. Future use of the presented lens model should thus acknowledge its limitations. 


\begin{figure*}
\centering          
\includegraphics[width=0.95\columnwidth,trim=0.6cm 0cm 1cm 0.8cm,clip]{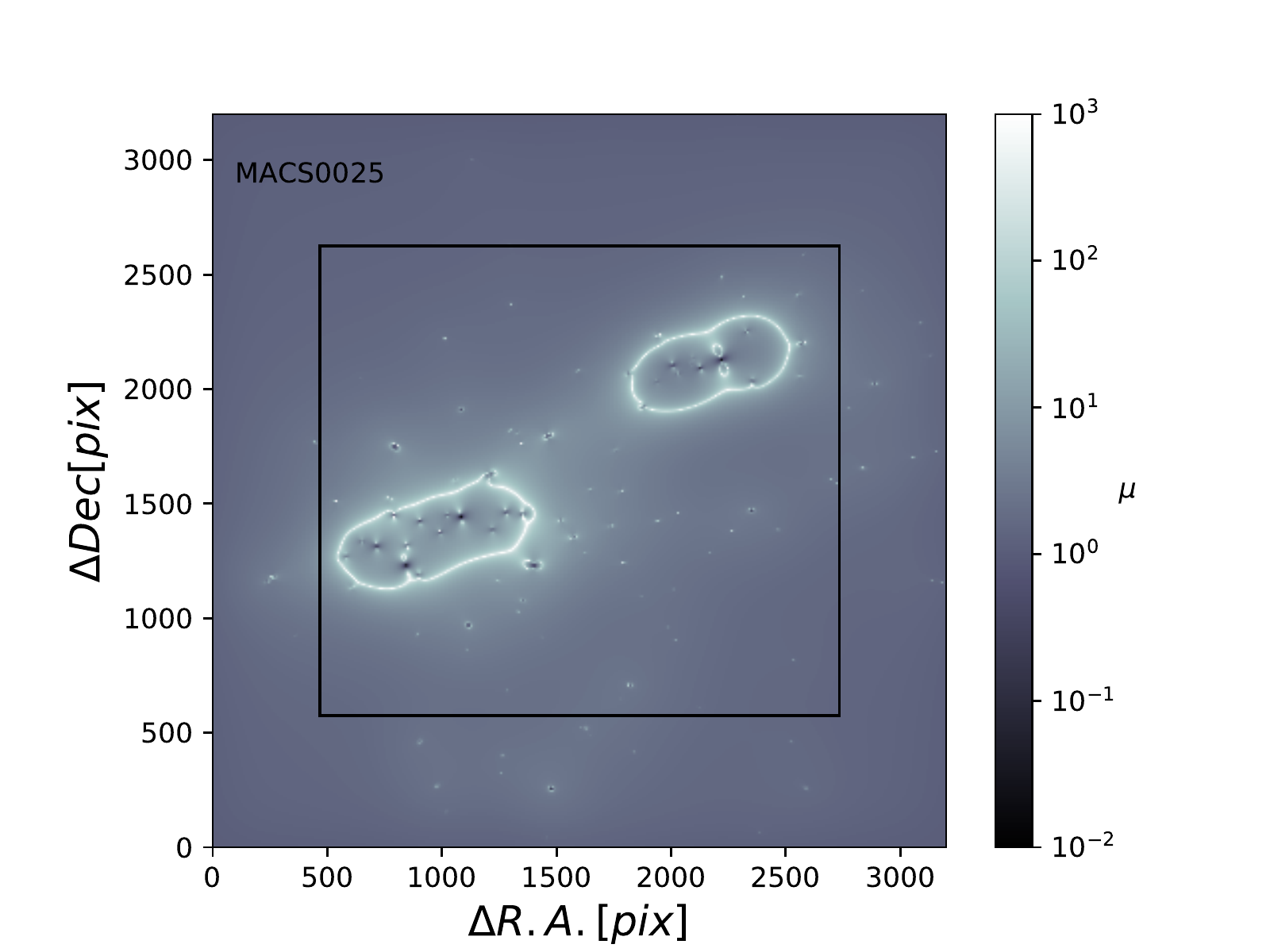}
\includegraphics[width=0.95\columnwidth,trim=0.6cm 0cm 1cm 0.8cm,clip]{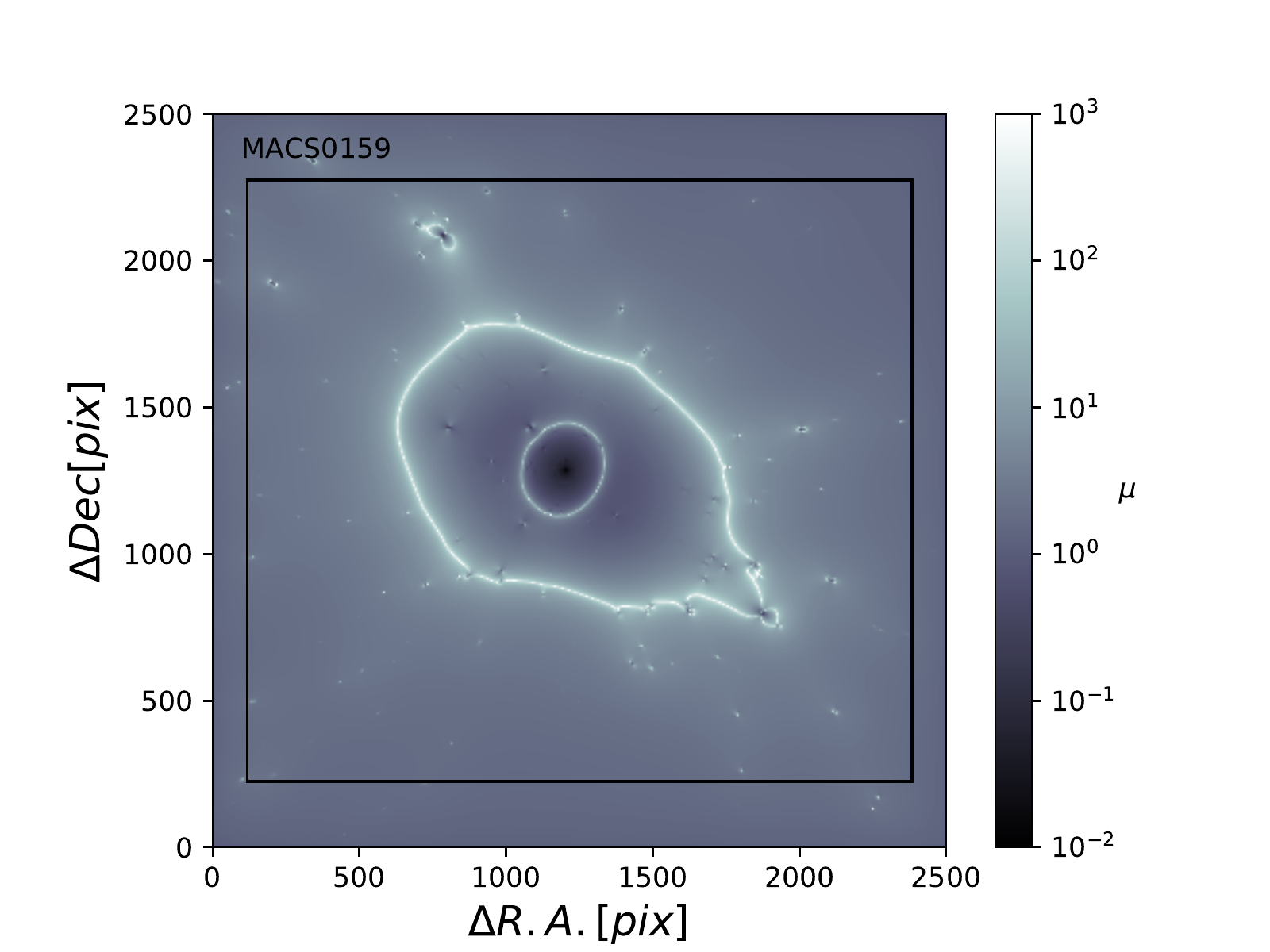}\\
\includegraphics[width=0.95\columnwidth,trim=0.6cm 0cm 1cm 0.8cm,clip]{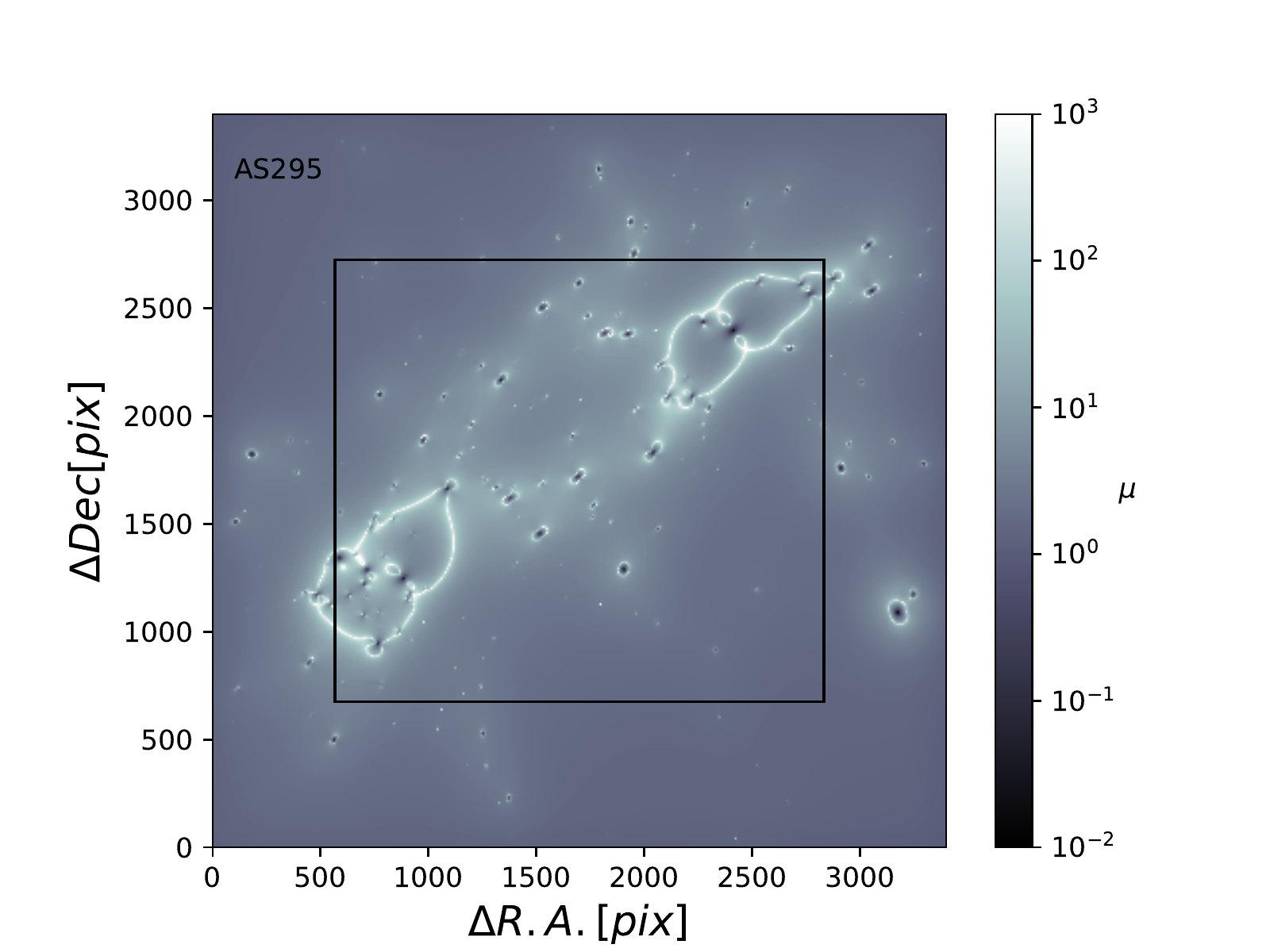}
\includegraphics[width=0.95\columnwidth,trim=0.6cm 0cm 1cm 0.8cm,clip]{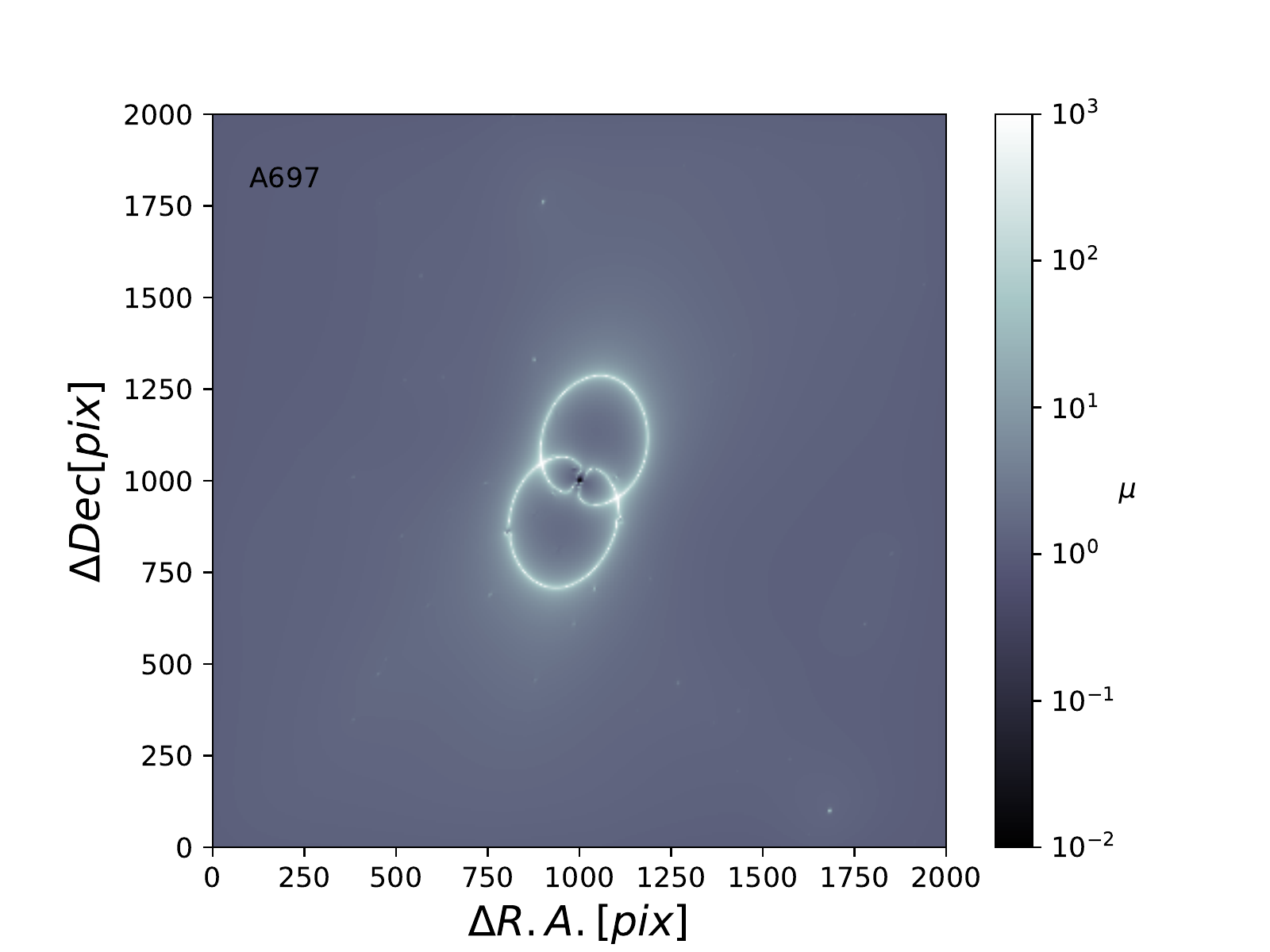}
\caption{Magnification maps for the four RELICS clusters analyzed in this work. The maps show the expected magnification distribution for a source located at $z = 9$, from our best-fit models. The orientations are the same as in Figures 1-4. The pixel scale is 0.06"/pixel. Black squares represent the WFC3/IR field-of-view ($136" \times 123"$). For A697 our FoV is fully encompassed within the WFC3/IR FoV.}
\label{mags}
\end{figure*}

\subsection{MACS J0159.8-0849}


We identify four multiple-image systems in MACS0159, with which we construct our model. The first and the second systems represent different constraints related to the same source (although it is unclear if the source is one or, e.g., two merging galaxies). The third system shows a distinctive green color -- in the color-composite RELICS HST image -- that aided in its identification, and the fourth system corresponds to a small, relatively bright source.

We set the main source to be at $z = 1.55$, corresponding to the mean z$_{\rm{phot}}$ from the counter images 1.1 and 1.2 (the more reliable BPZ estimates, since image 1.3 suffers contamination from the BCG light), and fix the redshift of systems 1 and 2 to this value. The redshifts of systems 3 and 4 are left as free parameters to be constrained by the model.

The mass distribution of the cluster is modeled with both the weight and the core radius of the BCG as free parameters, where the latter can reach values up to $100$ kpc. Our best-fit model does not include an ellipticity for the BCG, which seems to be fairly spherical. 

Our final model produces well all observed multiple images (Fig. \ref{stamp_0159}), resulting in an \textit{rms} of $\simeq 0.96"$. For each of systems 3 and 4 the model also predicts a third, considerably fainter image close to the cluster center, that we do not clearly identify in the data, likely due to the BCG's light and the expected faintness of these images. Fig. \ref{abellS295cc} shows the critical curves for our best-fit model, for a source at $z = 1.55$, and the position of the multiple images. 

\vspace{8mm}

\subsection{Abell S295}

The modeling of AS295 is based on the identification of six sets of constraints. The first four are part of a known giant arc located in the north-west region of the cluster, close to the BCG. These four subsets can be distinguished by the differences in their color and relative positions, where we identify three multiple images for each set. The arc has a spectroscopic redshift of $0.93$ reported in the literature \citep{edge1994}, and we adopt this value for all four subsets related to the arc. 

The two other systems (Systems 5 and 6) are found in the south-east concentration of galaxies. These are new identifications, with three multiple images each. System 5 has an elongated image on the east and two counter images lying close to each other, south of the BCG. Between these two images there is an additional faint galaxy that might locally contribute somewhat to the lensing as well. System 6 is a faint and long arc, north-west of the respective BCG. For these systems there is no spectroscopic redshift measurements available, so we leave their redshifts as free parameters of the model. We note that while we refer to systems 5 and 6 as secure here since they match the model's prediction, they lack clear internal details and thus their identification should be taken with slightly more caution.

Our model for AS295 has the weight of the two BCGs as free parameters to be optimized in the minimization. For the Northern BCG we set the ellipticity and position angle to the values given by \textsc{SExtractor}. The Southern BCG presents a spherical morphology and so we decided not to apply an ellipticity. The weight of two bright, spiral-looking cluster members (RA = 02:45:25.710, Dec = -53:01:43.031 and RA = 02:45:37.573, Dec=-53:02:58.595) lying near the two BCGs, was reduced by factor of 2 (this slightly improved the reproduction of multiple images; see also discussion in \S \ref{sec:dissLM} for MACS0025).

For systems 1 to 4 our model predicts well the three observed counter images in each subset. It also predicts a fourth, faint/small counter image close to the BCG for these systems. System 5 has a reasonable reproduction, and for system 6 we obtain a long arc predicted by the model, in agreement with the position and orientation of our image constraints. The total rms of the predicted images with respect to the observed locations is $\simeq  0.77 \arcsec$. The reproduction of multiple images can be seen in Fig. \ref{stamp_295}.

The resulting critical curves from our best-fit model (for a source redshift of $z = 1$), along with the multiple images, are seen in Fig. \ref{abellS295cc}. 



\subsection{Abell 697}

We construct the model for A697 using three multiple-image systems. Because there is no spectroscopic redshift available for any of the systems, we fix the  main source redshift to the value of the system with the most reliable photometric-redshift estimate (corresponding to system 1), leaving the redshift of the remaining systems as free parameters. 

The first system corresponds to a blue source showing a distinct morphology, leading to a reliable identification of four counter images well predicted by our model. The second system is also a blue-looking galaxy and we identify three counter images south and south-east of the BCG. The third is an arc-shaped object lying south of the cluster center, for which we identify two multiple images. This system was previously reported in the literature by \cite{metzger2000}, who has estimated a lower limit for the redshift of the source of $z > 1.3$ by combining the arc spectrum and its color. 

Our model for A697 allows the weight of the BCG to vary as a free parameter, as well as its ellipticity and position angle. The source redshift for system 1 is set to the mean of the $z_{\rm{phot}}$ distributions of images 1.2 and 1.4 (as image 1.1 suffers contamination from the BCG light and for image 1.3 there was no $z_{\rm{phot}}$ solution found). Redshifts for systems 2 and 3 are allowed to vary and are optimized by the minimization procedure. Our best-fit model yields a redshift of $z = 2.97_{-0.38}^{+0.04}$ for system 3, consistent with the lower limit found by \cite{metzger2000} within $1\sigma$. 

The final model reproduces well systems 1 and 2. For system 1 it also predicts a smaller fifth counter image very close to the BCG that we can not identify, possible due to the contamination by the BCG light. The model correctly predicts the arc-shape and orientation of system 3, and as expected from the lensing symmetry it also predicts a third, fainter counter image on the east side of the cluster center that we do not securely identify. The reproduction of multiple images by the best-fit model, which has a final rms of $\simeq 0.82 \arcsec$, is shown in Fig. \ref{stamp_697}. The resulting critical curves for a source at $z = 2$ and the multiple images used as constraints are shown in Fig. \ref{abell697cc}. 



\section{Results and Discussion} 
\label{sec:discuss}

\subsection{Lens modeling results}
\label{sec:dissLM}

Our best-fit model for MACS0025, as expected from the cluster galaxy distribution, exhibits a bimodal mass distribution (Figs. \ref{macs0025cc}, \ref{kappas}), in agreement with the findings by \cite{bradac08} and \cite{Zitrin2011}. Such morphology supports the picture that this cluster is undergoing a major merger \citep{bradac08,ma10}. The Einstein radius obtained from the modeling for the North-Western subcluster is $\theta_{\rm{E}}(z=2) = (8.1 \pm 0.8)"$ and encloses a mass of $M = (1.60 \pm 0.24) \times 10^{13} M_{\odot}$, while for the South-Eastern substructure we found $\theta_{\rm{E}}(z=2) = (7.2 \pm 0.7)"$ and enclosed mass $M = (1.18 \pm 0.18) \times 10^{13} M_{\odot}$. Note that \cite{Zitrin2011} reported a substantially larger radius for MACS0025, of $\theta_{\rm{E}} = (30 \pm 2)"$ for $z = 2.38$. Our model is similar to theirs around the North-Western subclump, but implies a much smaller critical curve around the main (South-Eastern) clump (as can be seen by comparing Fig. 3 from \citet{Zitrin2011} to our Fig. \ref{macs0025cc}). Our model was constructed with two major differences compared with the previous model by \citet{Zitrin2011}. First, as was implied by the RELICS data, here we assign a significantly higher redshift for System 2, so that the critical curves of the main mass clump are required to be smaller per given redshift. The second difference is the lower weight manually given by us to some of the bright galaxies around the South-Eastern main clump. Specifically, these are (apparently) cluster galaxies, that show bright spiral structure, and therefore seem to significantly deviate from the general, early-type cluster member M/L ratio, in the sense that they are about an order of magnitude too bright for their actual mass contribution compared to cluster ellipticals \citep[][]{Maraston2005,Courteau2014,Bahcall2014}. We therefore manually reduced the mass (i.e. input flux) of these galaxies by a factor of $10$. \footnote{Note, this estimate is subjective and based on our experience in a previous analysis of a large sample of clusters, e.g., \cite{Zitrin2015}. Nonetheless, once their weight has been substantially reduced, their exact mass contribution is not of particular importance.}

MACS0159 is a massive X-ray selected lens and has the largest (single-clump) Einstein radius among the clusters analyzed in this work, corresponding to $\theta_{\rm{E}}(z=2) = (24.9 \pm 2.5)"$. The mass inside the critical curves corresponds to $M = (1.15 \pm 0.17) \times 10^{14} M_{\odot}$. Our model exhibits a fairly round and regular morphology (Fig. \ref{kappas}), which together with the regular X-ray morphology \citep{maughan2008} might indicate that this is a relaxed cluster. The central flattening in the projected mass density profile seen in Fig. \ref{profs} (top-right), may be, in part, a result of the superposition of other elliptical cluster members onto the BCG (Fig. \ref{macs0159cc}).

The model for AS295 shows a bimodal mass distribution, following the cluster galaxy concentrations (Figs. \ref{abellS295cc}, \ref{kappas}). The North-Western clump hosting the giant arc has an Einstein radius of $\theta_{\rm{E}}(z=2) = (12.6 \pm 1.3)"$, and  mass of $M = (1.96 \pm 0.29) \times 10^{13} M_{\odot}$ enclosed by the critical curves. For the South-Eastern subcluster our best-fit model gives an Einstein radius of $\theta_{\rm{E}}(z=2) = (12.7 \pm 1.3)"$ and an enclosed mass of $M = (2.04 \pm 0.31) \times 10^{13} M_{\odot}$.  

\begin{figure}
	\includegraphics[width=1.1\linewidth]{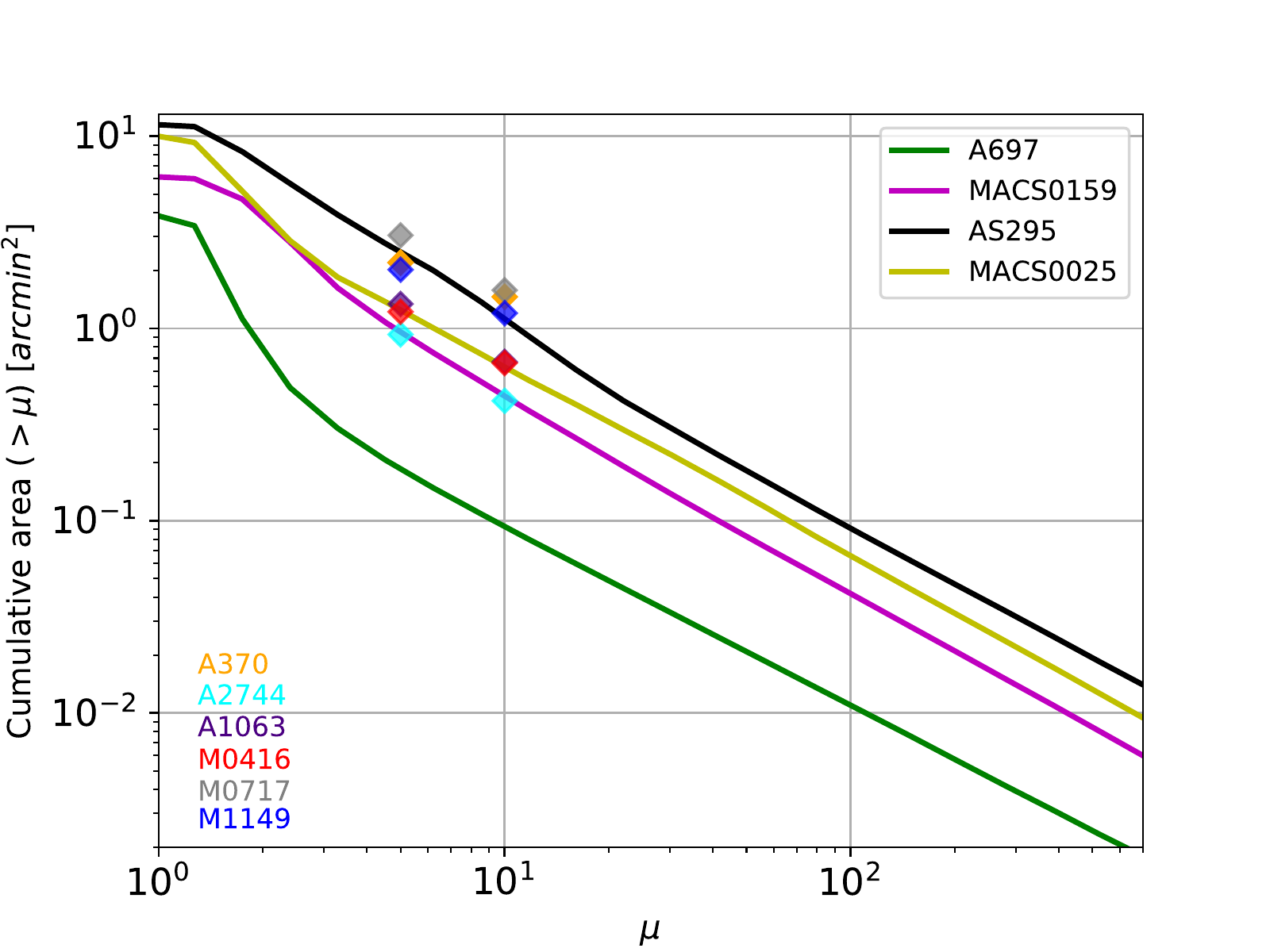} 
 	\caption{The image-plane area above each magnification value as a function of magnification value, for each of the analyzed clusters. The magnification is computed for a source redshift of $z = 9$. Diamonds indicate the cumulative area with $\mu \geq 5$ and $\mu \geq 10$ for the HFF clusters (considering the LTM-Gauss models).}
	\label{areaMag}
\end{figure}

Our analysis for A697 implies a rather elliptical mass distribution in projection (Fig. \ref{kappas}), and elongated critical curves (Fig. \ref{mags}). The Einstein radius given by the model is $\theta_{\rm{E}}(z=2) = (11.1 \pm 0.1)"$ and the critical curves encompass a total mass of $M = (1.45 \pm 0.23) \times 10^{13} M_{\odot}$. The somewhat-elliptical nature of the lens found in our analysis agrees with the model from \cite{metzger2000}, with the BCG orientation, and with the X-ray morphology reported in \cite{girardi2006}, which reinforces the possible recent-merger scenario. 

As a self-consistency check, we note that the Einstein radius and mass agree with the expectation given (for a circular lens) by:  

\begin{equation}
\theta_E=\left(\frac{4GM(<\theta_E)}{c^2} \frac{D_{ls}}{D_l D_s}\right)^{1/2}.
\end{equation}



\begin{figure*}
\centering
\includegraphics[width=.42\linewidth]{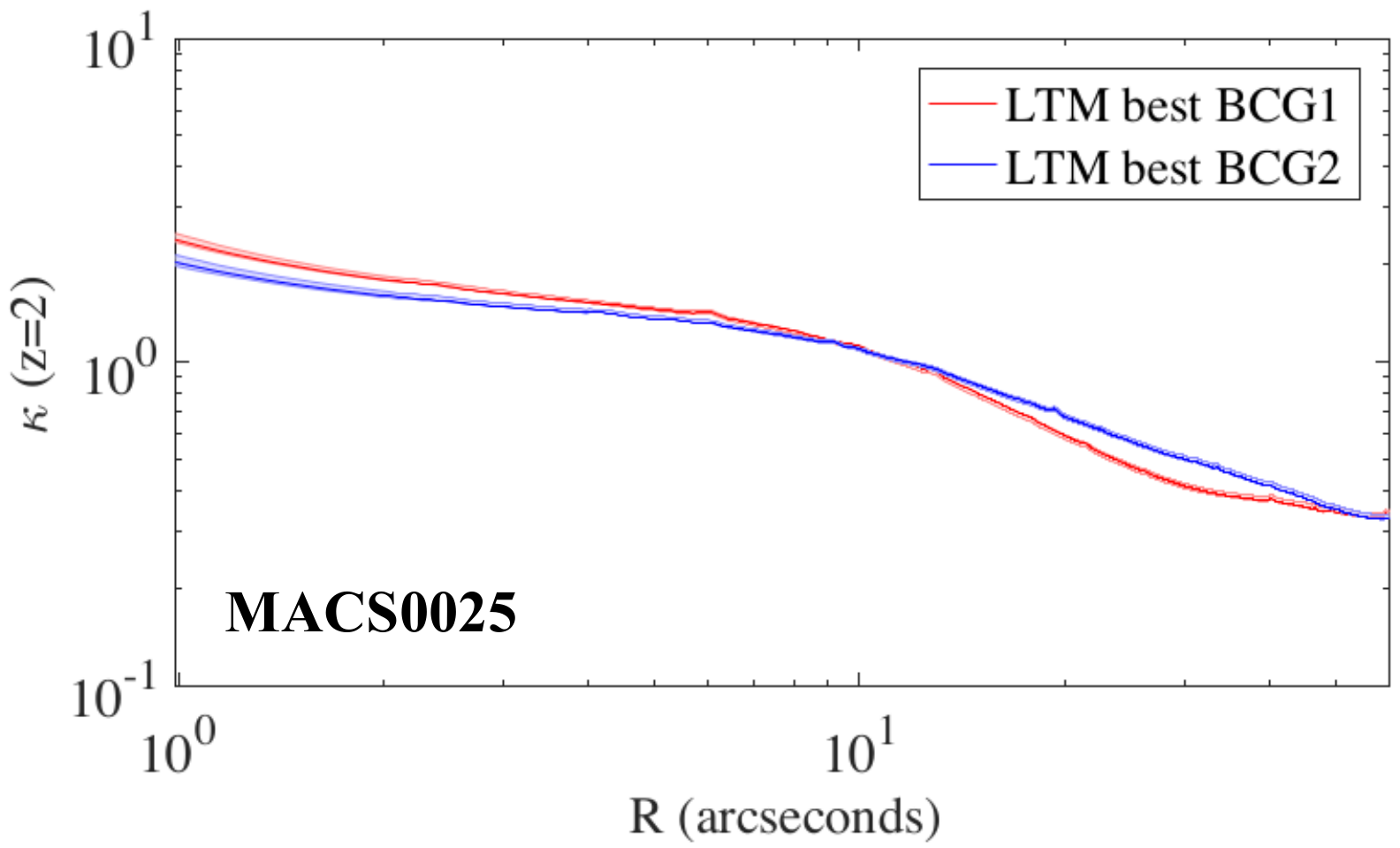}
	\includegraphics[width=0.42\linewidth]{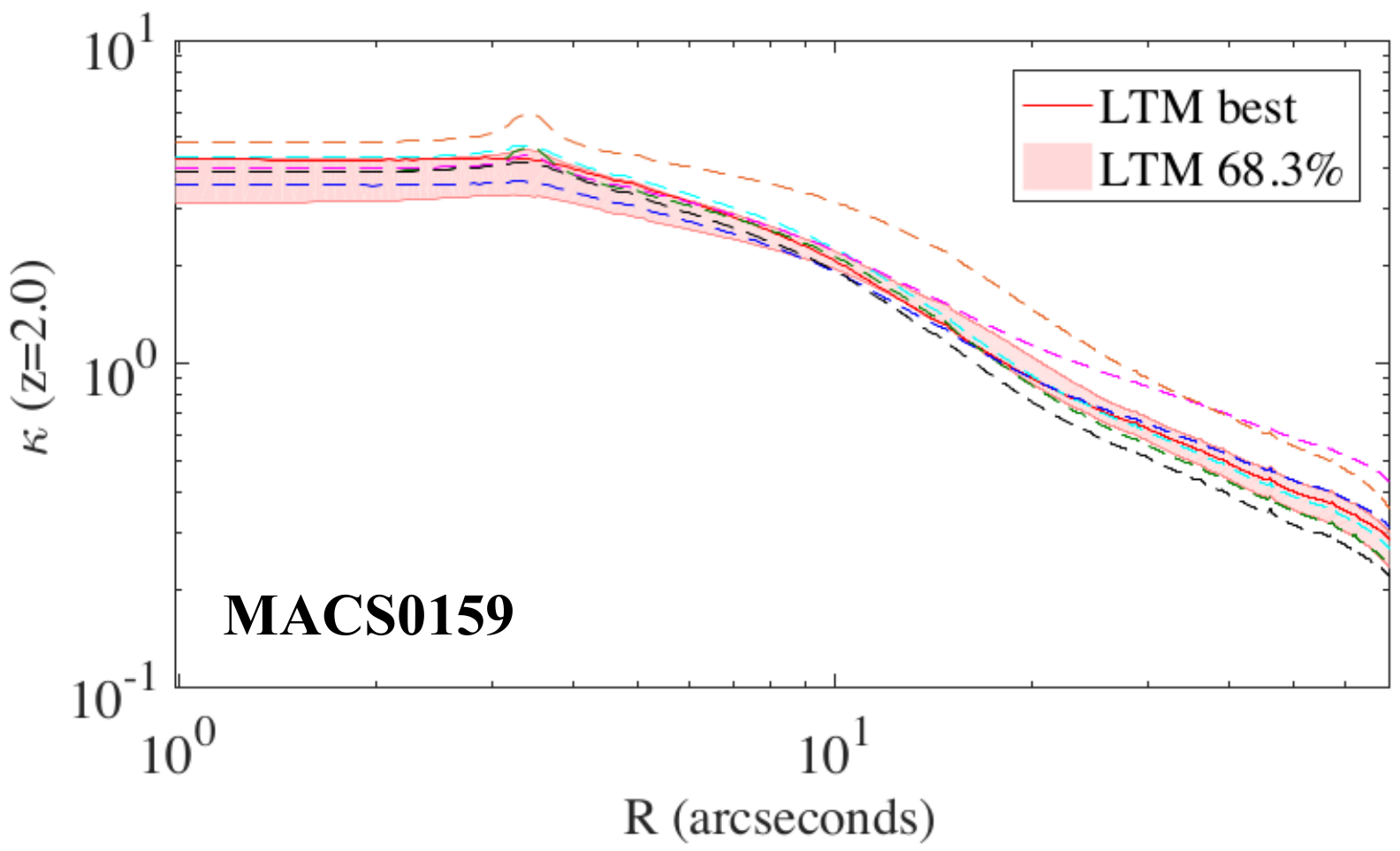} \\
\includegraphics[width=.42\linewidth]{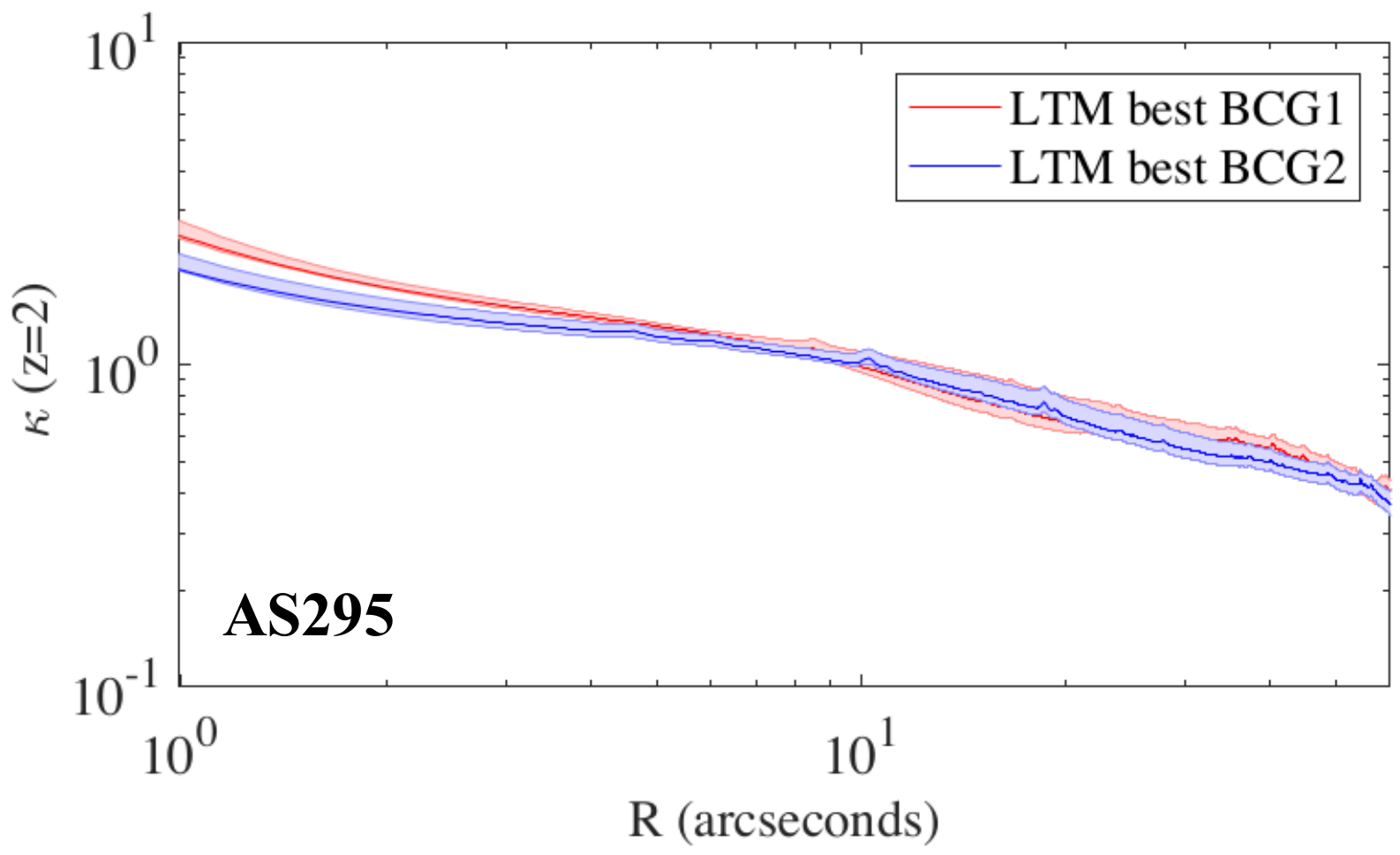} 
    \includegraphics[width=0.42\linewidth]{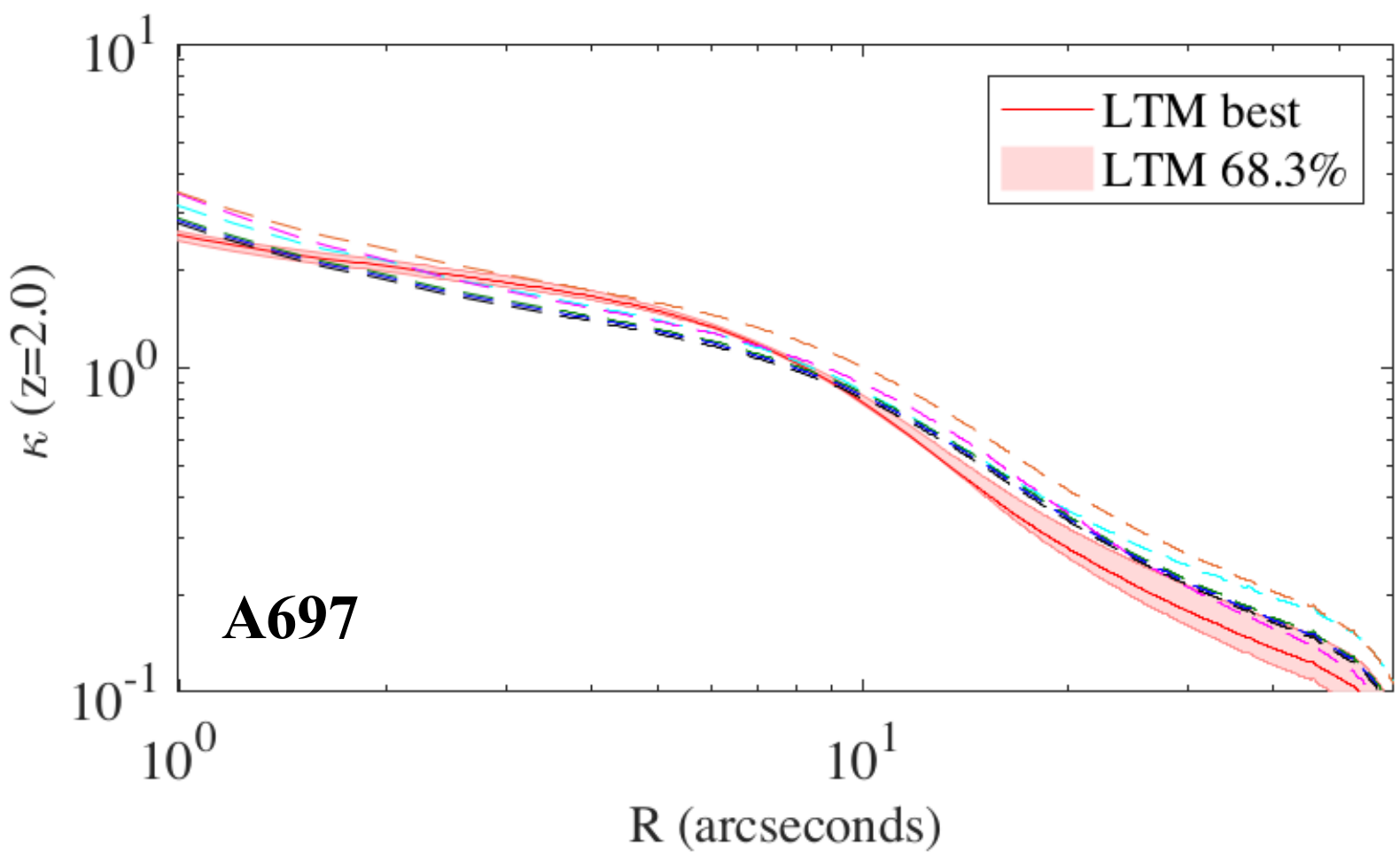}
 	\caption{Radial mass-density profiles in terms of the convergence $\kappa$, scaled to $z = 2$, for the RELICS clusters analyzed in this work. For MAC0025 and AS295 we show the profile around the BCG of each subcluster, namely BCG1 for the Northern-Western clump and BCG2 for the South-Eastern clump. For the clusters lacking of $z_{\rm{spec}}$, MACS0159 and A697, we also present profiles computed by assuming different redshifts for the main system together with the result of the best-fit model, shown by the solid red line together with the $68.3\%$ confidence level represented by the shaded red region. Cyan and green lines corresponds to a change of $-10\%, +10\%$ in redshift, respectively, magenta and blue lines to $-25\%, +25\%$ and orange and black to a variation of $-50\%, +50\%$ respectively.}
	\label{profs}
\end{figure*} 

\subsection{Implications for high-redshift studies}
\subsubsection{Lensing strength}

Galaxy clusters with large Einstein radii and those that entail a large area of high magnification are objects of particular interest in the search for high-z galaxies \citep[e.g.,][]{Zheng2012,Zitrin2014,Kawamata2016}. One way to evaluate the lensing strength of a lens is through the total area, A, in the source plane or the image plane, that has a magnification above a certain value, as a function of the magnification value (a plot of $A(>\mu)$ versus $\mu$). This area is sometimes referred to as the cumulative area above a certain magnification \citep[e.g.,][]{Cerny2017}, a term we use here as well. In Fig. \ref{areaMag} we show such cumulative magnification curves computed from our best-fit models, for a source located at redshift $z_{s}=9$. We also present therein the cumulative areas of the HFF clusters, specifically, for $\mu \geq 5$ and $\mu \geq 10$ (adopting the Zitrin-LTM-Gauss models, as these were constructed with similar formalism as the one used here). 

Among the four clusters, AS295 shows the largest high magnification area, almost comparable to the largest HFF clusters, as can be seen in Fig. \ref{areaMag} \citep[see also][]{Johnson2014}. However, given the lack of spectroscopic redshifts in our current analysis, this comparison should be taken with caution. As detailed in the next section, we expect a typical variation of $\pm 50\%$ over the cumulative area of high magnification arising from the redshift uncertainties. MACS0159 and MACS0025 seem to also be quite strong lenses in terms of cumulative magnification area, as strong as some of the HFF clusters. A697 is the smallest lens in our analysis, showing the smallest cumulative area above any magnification value (Fig. \ref{areaMag}).

The fact that the most regular- or relaxed-looking cluster in our analysis, A697, shows a relatively small Einstein radius, and in addition a smaller area of high magnifications, is not surprising. As can be seen for example in Fig. \ref{mags}, merging clusters create an area of high magnification between the merging subclumps (in addition to, or because of the lensing strength of each subclump individually). This means that although the total critical area (summed over the subclumps) can be modest or even small, merging clusters, projected on the plane of the sky, will often show large areas of high magnification. The fact that the two largest cumulative areas of high magnification found in our analysis correspond to the two clusters showing substructures demonstrates this idea. Note that the HFF clusters are also, similarly, merging clusters. The impact of cluster mergers on lensing properties has been examined before. It has been found that merging clumps boost the critical area - and the number of multiple images in case the subclumps are close enough - forming the most useful strong gravitational lenses \citep[see e.g.,][]{Torri2004,Meneghetti2007,Redlich2012,Zitrin2013}, especially if also the critical curves are relatively large as is the case for the HFF clusters. In that sense, the lensing strength estimate by the cumulative area alone might be misleading. Merging clusters, such as those of the HFF which have both large enough critical areas and also large high magnification areas, are preferable at least for two reasons: first, they will be rich in multiple images and are thus much better modeled. Second, if the subclumps are too small or distant so that the critical curves do not merge and the critical area remains effectively small -- such as for MACS0025 and AS295 here -- the area of high magnification, even if large, will also be highly sheared, so that background galaxies will be significantly stretched perpendicular to the axis connecting the two subclumps. It has been found that high shear can significantly hinder the detection of magnified high-redshift galaxies \citep{Oesch2015,Bouwens2017a}. If the critical area is large enough, multiple images of the background sources will typically appear, and so the combination of a large high magnification area and large enough critical area seems crucial for maximizing the detection of intrinsically faint high-redshift objects. In that sense, we conclude that the intrinsically ``best lens" in our four cluster sample, is likely MACS0159.

We also note that the cumulative-area comparison (Fig. \ref{areaMag}) to estimate the lensing strength is sometimes more insightful if the same field-of-view (FOV) is probed in all clusters (for example it is often comfortable to define the WFC3/IR FOV, in the case of HST imaging), especially if the goal is to plan observations and estimate the expected number of objects per FOV. Here, however, our goal is to maximize the area probed for each cluster and so we simply consider the full FOV used for the modeling, which differ somewhat from one another (the $A(\mu=0)$ points in Fig. \ref{areaMag} represent the total area modeled for each cluster). The somewhat different FOV sizes, however, mostly affect the areas with low magnifications, as high magnifications are generally induced only near the center and are thus included in any case (increasing the FOV will shift the $A(\mu=0)$ point but will hardly affect the $A(>\mu=10)$ point, for example).

Another related important point is why A697 is a relatively small lens, given it is the tenth most massive SZ-mass cluster in the Planck catalog \citep{Planck2016}. While massive, often merging, bright X-ray clusters \citep{Ebeling2001} have proven to be exceptional strong lenses \citep[e.g.,][and references therein]{Zitrin2013}, it should be examined more statistically if the scatter on SZ-mass, usually estimated through X-ray or weak-lensing mass scaling relations, is an equal indicator for SL strength. It is also possible that the SZ signal in this case was boosted by an underlying radio source \citep{kempner2001,Venturi2008,macario2010}. We leave these issues to be examined in other, future work.

\begin{table*}
	\caption{High-z ($z \sim 6$) candidates}            
	\label{table:highzcan}      
	\centering  
    {\renewcommand{\arraystretch}{1.5}
	\begin{tabular}{c c c c c c c c c}        
		\hline\hline                 
		Galaxy ID\tablenotemark{a} & R.A. & Dec& $J_{125}$\tablenotemark{b} & $z_{\rm{phot}}^{EAZY}$ & $z_{\rm{phot}}^{BPZ}$& $\mu_{\rm{best}}$\tablenotemark{c} & $\mu$\tablenotemark{d} &  M$_{\rm{UV,}1500}$\tablenotemark{e} \\  
		&(J2000)&(J2000)& [AB] &&&&& [AB]\\
		\hline 
  MACS0025-12-0169 & 00:25:29.89 & -12:22:12.76 & $25.96 \pm 0.24$ & $6.0^{+0.1}_{-5.6}$ & $5.4^{+0.5}_{-5.0}$ & 2.28 & $2.31^{+0.03}_{-0.04}$ & $-20.00^{+0.30}_{-0.32}$ \\  
  MACS0025-12-0450 & 00:25:32.74 & -12:22:40.02 & $25.80 \pm 0.22$ & $5.9^{+0.0}_{-5.4}$ & $5.3^{+0.5}_{-4.8}$ & 2.17 & $2.23^{+0.03}_{-0.05}$ & $-20.41^{+0.30}_{-0.32}$ \\    
  MACS0025-12-0554 & 00:25:25.01 & -12:22:51.38 & $25.79 \pm 0.22$ & $6.5^{+0.3}_{-4.8}$ & $6.3^{+0.3}_{-0.5}$ & 1.38 & $1.40^{+0.01}_{-0.01}$ &  $-20.62^{+0.30}_{-0.30}$ \\  
  MACS0025-12-0770 & 00:25:33.92 & -12:23:08.83 & $26.82 \pm 0.36$ & $5.7^{+0.2}_{-4.7}$ & $0.7^{+4.7}_{-0.2}$ & 2.07 & $2.13^{+0.03}_{-0.05}$ & $-19.05^{+0.30}_{-0.33}$ \\   
  MACS0025-12-0851 & 00:25:35.76 & -12:23:14.03 & $26.90 \pm 0.32$ & $5.9^{+0.1}_{-5.6}$ & $0.8^{+5.0}_{-0.3}$ & 1.88 & $1.91^{+0.01}_{-0.02}$ & $-18.94^{+0.30}_{-0.31}$ \\   
  MACS0025-12-1037 & 00:25:30.79 & -12:23:34.51 & $27.23 \pm 0.38$ & $6.4^{+0.5}_{-5.7}$ & $1.3^{+5.3}_{-0.6}$ & 11.58 & $12.48^{+0.68}
_{-0.79}$ & $-16.97^{+0.36}_{-0.55}$ \\
  MACS0025-12-0748 & 00:25:34.01 & -12:23:05.87 & $27.17 \pm 0.34$ & $6.7^{+1.1}_{-5.9}$ & $6.4^{+0.9}_{-5.5}$ & 2.11 & $2.12^{+0.03}_{-0.05}$ & $-18.67^{+0.30}_{-0.32}$ \\
  \hline
  MACS0159-08-0085 & 01:59:48.71 & -08:48:58.43 & $27.01 \pm 0.40$ & $6.0^{+0.2}_{-5.2}$ & $5.4^{+0.5}_{-4.7}$ & 2.10 & $2.06^{+0.20}_{-0.17}$ & $-19.32^{+0.32}_{-0.37}$ \\   
  MACS0159-08-0137 & 01:59:47.69 & -08:49:08.00 & $27.04 \pm 0.37$ & $6.3^{+0.4}_{-5.4}$ & $6.0^{+0.5}_{-5.2}$ & 2.15 & $2.15^{+0.23}_{-0.18}$ & $-19.06^{+0.32}_{-0.37}$ \\   
  MACS0159-08-0661 & 01:59:47.18 & -08:50:14.18 & $26.58 \pm 0.25$ & $5.8^{+0.2}_{-0.7}$ & $5.6^{+0.3}_{-0.7}$ & 14.63 & $14.85^{+3.95}_{-5.12}$ & $-17.24^{+0.55}_{-1.22}$ \\ 
  MACS0159-08-0621 & 01:59:50.29 & -08:50:08.05 & $27.96 \pm 0.49$ & $7.1^{+1.1}_{-5.6}$ & $6.6^{+1.0}_{-5.5}$ & 5.62 & $5.69^{+0.79}_{-0.72}$ & $-17.30^{+0.36}_{-0.54}$ \\
  \hline
  Abells295-0250 & 02:45:29.87 & -53:01:50.40 & $24.93 \pm 0.11$ & $6.3^{+0.3}_{-0.2}$ & $6.3^{+0.3}_{-0.2}$ & 2.60 & $2.70^{+0.51}_{-0.34}$ & $-20.68^{+0.32}_{-0.31}$ \\
  Abells295-0355 & 02:45:27.92 & -53:02:00.14 & $27.76 \pm 0.50$ & $6.2^{+0.4}_{-5.6}$ & $5.9^{+0.4}_{-5.4}$ & 2.66 & $2.83^{+0.73}_{-0.61}$ & $-18.33^{+0.33}_{-0.47}$ \\   
  Abells295-0796 & 02:45:27.26 & -53:02:49.85 & $28.23 \pm 0.67$ & $6.1^{+0.3}_{-5.1}$ & $1.0^{+5.1}_{-0.4}$ & 6.60 & $6.82^{+1.41}_{-1.25}$ & $-17.14^{+0.35}_{-0.46}$ \\   
  Abells295-1055 & 02:45:25.15 & -53:03:18.86 & $25.52 \pm 0.17$ & $6.0^{+0.4}_{-0.5}$ & $5.9^{+0.3}_{-0.3}$ & 2.43 & $2.43^{+0.28}_{-0.17}$ & $-20.13^{+0.31}_{-0.32}$ \\   
  Abells295-0737 & 02:45:30.05 & -53:02:43.65 & $27.13 \pm 0.29$ & $7.4^{+0.8}_{-6.4}$ & $6.5^{+1.2}_{-5.5}$ & 12.66 & $16.02^{+6.29}_{-6.23}$ & $-16.48^{+0.48}_{-0.49}$ \\ 
  Abells295-0568 & 02:45:36.25 & -53:02:25.87 & $26.02 \pm 0.17$ & $8.1^{+0.3}_{-1.7}$ & $7.7^{+0.4}_{-0.9}$ & 1.74 & $1.77^{+0.22}_{-0.13}$ & $-20.20^{+0.31}_{-0.36}$ \\
  \hline
  Abell697-0095 & 08:42:58.55 & 36:22:46.46 & $26.12 \pm 0.24$ & $5.7^{+0.2}_{-0.8}$ & $5.5^{+0.3}_{-0.5}$ & 1.24 & $1.28^{+0.01}_{-0.01}$ & $-20.62^{+0.30}_{-0.33}$ \\    
  Abell697-0184 & 08:42:57.33 & 36:22:19.10 & $26.41 \pm 0.28$ & $6.1^{+0.3}_{-5.3}$ & $5.7^{+0.4}_{-5.0}$ & 1.70 & $1.79^{+0.05}_{-0.05}$ & $-19.96^{+0.31}_{-0.47}$ \\    
  Abell697-0636\tablenotemark{f} & 08:43:01.24 & 36:21:35.75 & $25.69 \pm 0.18$ & $6.0^{+0.6}_{-5.2}$ & $6.0^{+0.5}_{-5.1}$ & 1.49 & $1.55^{+0.03}_{-0.02}$ & $-20.31^{+0.31}_{-0.46}$ \\   
  Abell697-0972\tablenotemark{g} & 08:43:00.14 & 36:20:58.02 & $26.98 \pm 0.31$ & $6.3^{+0.8}_{-6.0}$ & $0.8^{+5.4}_{-0.4}$ &- &-&- \\
		\hline\hline                           
	\end{tabular}}
	\tablenotetext{1}{Following \citet{Salmon2017} notations.}
    \tablenotetext{2}{Apparent magnitude in the F125W band.}
    \tablenotetext{3}{Magnification from the best-fit model.}
    \tablenotetext{4}{Mean magnification and $1\sigma$ errors from 100 random MCMC realizations.}
     \tablenotetext{5}{Absolute magnitude at $\lambda = 1500$ \AA. Errors include the uncertainty in the fit to the UV slope and propagated photometric and magnification errors.}
         \tablenotetext{6}{z$_{spec} = 5.800$, obtained from our GMOS spectroscopic observations for A697.}
    \tablenotetext{7}{Outside the modeled FOV.}
    \label{tab:cand_z}
\end{table*}

\begin{figure*}
\centering          
\includegraphics[width=1.5\columnwidth]{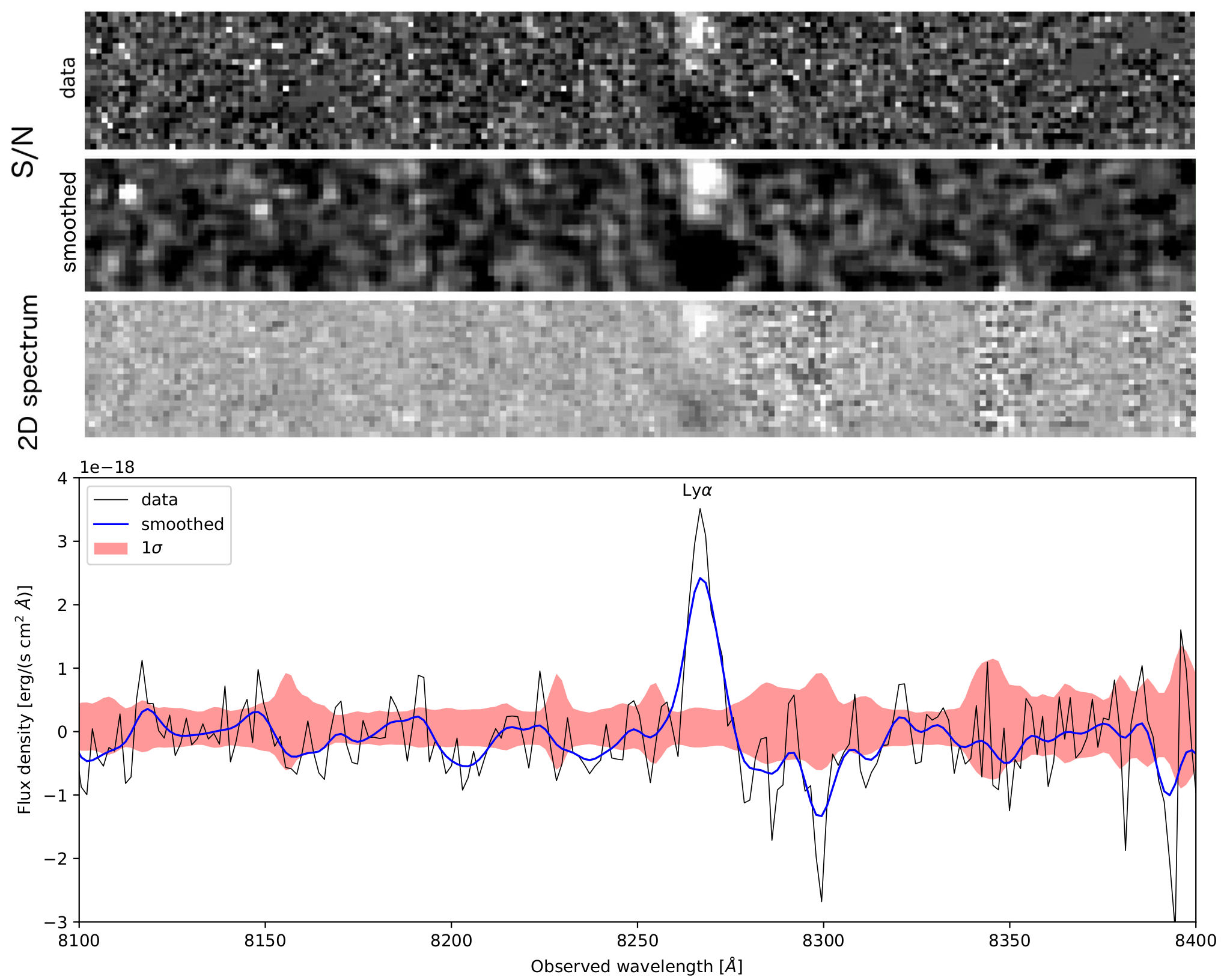}
\caption{Spectroscopic detection of an emission line in the high-z candidate Abell697-0636. The upper panels show the original and smoothed 2D signal-to-noise map, and the 2D spectrum around the detected line. Positive fluxes are indicated in white. In the lower panel we present the 1D data (solid black line) and smoothed (solid blue line) box-extracted 1D spectrum showing the emission line, together with the $1\sigma$ error (red regions), at the observed wavelength. The detection, interpreted as Ly$\alpha$, yields to $z = 5.800 \pm 0.001$. Another possible interpretation is [O II] at $z \sim 1.2$.} 
\label{spec7}
\end{figure*}

\subsubsection{RELICS high-redshift candidates}
The first sample of high-z galaxy candidates detected in RELICS clusters was recently presented by \cite{Salmon2017}. Regarding the clusters analyzed in this work, \cite{Salmon2017} found six candidates in MACS0025, three in MACS0159, four in AS295 and four in A697, in the redshift range $6 \leq z_{phot} \leq 8$. The photometric redshifts are estimated with both BPZ and Easy and Accurate Z \citep[EAZY,][]{Brammer2008}. The candidates are presented in Table \ref{tab:cand_z}, following the notation from \cite{Salmon2017}. Here, we provide magnification estimates for the candidates. In table \ref{tab:cand_z} we quote the best-fit magnification from our models, (corresponding to the minimum $\chi^2$), the average magnification from the MCMC, and the $68.3\%$ confidence interval obtained from a 100 random realizations from the MCMC fitting. Absolute magnitudes $M_{\rm{uv}}$ at the rest-frame $\lambda_{\rm{r}} = 1500$ \AA\ are computed for the candidates based on a fit to the UV continuum slope using the WFC3/IR bands (F105W, F125W, F140W, F160W), assuming the flux follows a power-law relation $f_{\lambda} \propto \lambda^{\beta}$ \citep[e.g.,][]{Meurer1999}. The final absolute magnitude is then obtained from the flux at $(1+z) \lambda_{\rm{r}}$.

Based on our models we also checked for -- but did not find -- any potential multiple image systems consistent with the reported high-redshift candidates among the analyzed clusters.

As detailed in \S \ref{gmos}, we spectroscopically observed the field of A697 using GMOS for 64.5 minutes. While no multiple image redshifts were obtained, we identified a single prominent ($\sim 15\sigma$) emission line in the high-z candidate Abell697-0636 from \citet{Salmon2017}. In Fig. \ref{spec7} we show the one and two-dimensional spectra, along with the signal to noise map.
Given our primary goal was to simply identify redshifts for multiple images, no spectrophotometric standards were observed. To estimate the line flux, we made use of the three alignment stars observed on our mask. The calibration was derived based on the spectroscopic data of the stars around the wavelength where the line is detected and the photometric measurements in the F814W filter. Two out of the three stars yield to calibrations in agreement within $3\%$; the third star gives a calibration which is lower by $\sim 35\%$. We adopt the average of the three and, based on their differences, assume a conservative calibration uncertainty of $\pm35\%$. This also encompasses an independent calibration we perform using a standard star observed with GMOS as a baseline calibration two weeks prior to our observations. The total line flux measured is $(1.5 \pm 0.5) \times 10^{-17}$ erg s$^{-1}$ cm$^{-2}$.

Interpreting the line as Ly$\alpha$ results in a rest-frame equivalent width (EW$_0$) of $(52 \pm 22)$ \AA. This is a common value for Ly$\alpha$ emitting galaxies at $z\sim6$ \citep{Ouchi2008,Kashikawa2011}. From a Gaussian fit to the line we derived a width of (117 $\pm$ 15) km s$^{-1}$ and observed FWHM of (7.6 $\pm$ 1) \AA, with the peak of the line centered at ($8267.2 \pm 0.4$) \AA, placing the object at a redshift of 5.800$\pm$ 0.001.

Other scenarios that could result in a single spectral feature correspond to a lower redshift galaxy. Considering the case of [O III] $\lambda\lambda$4959, 5007 or H$\beta$ $\lambda$ 4862, would yield $z \sim 0.7$. However, given the typical line flux ratios, we would expect the detection of at least one more line, which is not seen. Another possible interpretation is [O II] $\lambda\lambda$3726,3729 at $z \sim 1.2$. The [O II] scenario implies a rest-frame EW$_0$ of 178$^{+25}_{-43}$ \AA, which is higher than typical values ($\leq 100$ \AA) \cite[e.g.][]{Hogg1998,Pirzkal2013,Khostovan2016}, even though rare EW$_0$ exceeding this limit have been reported \citep{Stern2000}. Additionally, the (single) line width we measure seems to be too narrow to accommodate the typical peak separation of the [O II] doublet ($\sim 220$ km s$^{-1}$, \citealt{Kirby2007,Matthee2018}), further strengthening the Ly$\alpha$ interpretation.

Another aspect to be considered when interpreting the line detection is the observed colors of the host galaxy. In the case of Abell697-0636, there is no detection in the ACS F435W and F606W filters, while the flux increases towards the WFC3/IR filters. Based on the EW$_0$ values, line width, the observed dropout nature (and thus red color) and the SED fit from both BPZ and EAZY, we suggest -- although we cannot rule out the [O II] scenario completely -- that the detection most likely arises from Ly$\alpha$ emission. 

\subsection{Redshift and other uncertainties}\label{line}

A critical point in the construction of a SL model is the knowledge of the relative cosmological distances between observer, lens and source. The lack of spectroscopic information for the lensed galaxies may bias the final mass distribution and magnification maps derived from SL constraints \citep{Smith2009,Johnson2014,Johnson2016}. Two of our clusters, MACS0159 and A697, do not have any spectroscopic information available for the multiple-image systems, and we fixed the redshift of the source with the most reliable photo-z estimate (we often call this the "main source", the source the model is primarily calibrated to). The other two clusters, MACS0025 and AS295, have one system each with spectroscopic redshift available from the literature \citep{bradac08,edge1994}, to which our models are calibrated. 

Aside from MACS0025, for which we built a simple model based on two multiple-images systems with both redshifts fixed, for the other clusters we only fix the main source redshift and leave the redshift of the other systems to be constrained in the minimization, adopting a broad prior range (typically $\pm0.3$ in the relative $D_{\rm{ls}}/D_{\rm{s}}$ ratio). Given the uncertainties in source redshifts, it is essential to examine how such uncertainties affect our final solutions, especially for the two clusters that lack any spectroscopic redshifts (MACS0159 and A697). For these two clusters we conducted a series of tests where we shifted the main source redshift by $\pm 10\%$, $\pm 25\%$ and $\pm 50\%$ from its original photo-z value. The resulting mass density profile for each model is shown in Fig. \ref{profs} (right panels), together with the best-fit and $1\sigma$ interval of our fiducial models. Despite the expected variation in the profiles, the results show overall an agreement with our best-fit models within $3\sigma$ in most of the radial range. Differences exceeding the $3\sigma$ level for A697 occur at the very central region and at intermediate radius (around $\sim 8"$), where the errors on the best-fit profile are particularly small. For MACS0159 the extreme case of a $-50 \%$ shift results in a higher profile, above the $3\sigma$ level in all scales. 

We use the same suite of trial models to examine the effect of lack of spectroscopic redshifts also on the power of the lens (i.e., the cumulative magnification area). We find that the redshift uncertainties typically propagate into a difference of up to $\pm 50\%$ percent in the cumulative area of the lens with high magnification (for example, above $\mu>5$ or $\mu>10$). The exception, as expected from the density profile results, is the $- 50\%$ change in redshift. In this case the area of high magnification can be increased by up to $100\%$ for both MACS0159 and A697. We note, however, that this most extreme effect increases the lensing strength, indicating that our magnification estimate is conservative. The impact of redshift errors is generally smaller in areas with lower magnification values. In a similar fashion, for example, \citet{Cerny2017} found for several clusters that about 90\% of their modeled FOV ($200\arcsec \times 200\arcsec$) was typically constrained to better than $\sim 20-40\%$.

To similarly examine the choice of redshift constraint configurations on the models of MACS0025 and AS295, we also run models changing the initial set-up. Explicitly, for MACS0025 we run two models, one in which we leave the redshift of system 2 free, and another in which the redshift of system 2 is fixed to its photometric redshift value, while the redshift of system 1 is kept fixed. The resulting mass profiles are shown in Figure \ref{0025profs}. Both new models lead to results for the North-Western subclump fairly similar to the fiducial one. Especially, the redshift of system 1 in the case where it was left free, is $\sim2.6$, not substantially different than its spectroscopic redshift (2.38). For the South-Eastern subclump, the Einstein radius and mass encompassed by the critical curves can increase by up to $80\%$.  

For AS295, we run a model in which we fix the redshift of system 6 to the respective photo-z value, while leaving systems 1-5 free. From this trial we also want to test if we are able to recover the redshift for systems 1-4. A comparison between the fiducial and the new mass profiles is shown in Figure \ref{295profs}. The new model predicts the redshift of systems 1-4 within $\pm 0.05$ of the spectroscopic value ($z_{spec} = 0.93$) fixed in the original model. Differences between the Einstein radius and mass enclosed by the critical curves, for both subclumps, are in the range of few percents ($4 - 8\%$).

With respect to the area of high magnification, we found that the largest deviation is around $\pm 30\%$ for both MACS0025 and AS295.

Another aspect that should be noted is the model redshift prediction for multiple image systems whose redshifts were left to be optimized by the model, compared to their photometric redshift. This is of particular importance as in few cases the model redshift predictions, seen in Tables \ref{table:0025}-\ref{table:697}, seem to differ significantly from the photometric redshifts, and so we wish to test whether this has a substantial effect on the resulting models. 

For that reason, in addition to the tests mentioned above, we also run models for MACS0159 and A697 in which we set the redshift of systems that were initially left free to their best-fit photometric values, and compare to the original results. For cases in which different images from the same system have somewhat different photo-z values, we left the redshift of these systems to vary within the range of such estimates. The profiles arising from this experiment are shown in Figure \ref{0159_697profs}.

We find that the new model for A697 gives results in close agreement with the original model, both in terms of Einstein radius and mass encompassed by the critical curves as well as cumulative area of high magnification ($A(> \mu = 5)$ and $A(> \mu = 10)$). In general the results agree within few percents ($< 5\%$).

For the new tested model of MACS0159, we recover an Einstein radius and corresponding mass that are $\approx 13\%$ smaller than the fiducial ones. This model also results in an area of high magnification lower by $~ 35\%$ (for both A($> \mu = 5$) and A($> \mu = 10$)) compared to the initial model.

Other sources of systematic uncertainties in SL models include the non-correlated matter along the line-of-sight \citep{Seljak1994,Puchwein2009,Daloisio2014,Bayliss2014}, the choice of parametrization \citep{Zitrin2015,Meneghetti2017}, identification of false multiple-image systems \citep{bradac08}, the mass-sheet degeneracy \citep{Falco1985,Bradac2004,Liesenborgs2012} and uncertainties in the adopted cosmology \citep{Bayliss2015}. The understanding of systematics is central for studies based on strong lens models such as measurements of the high-z luminosity function, which relies on the determination of intrinsic properties of lensed galaxies and of the effective volume. Such uncertainties should be taken into account when propagating results based on SL models, and we refer the reader to the above works for further discussion on these. 

Regardless of the above uncertainties, we additionally emphasize that at the outskirts of the model FOV, beyond the area where constraints from multiple-images are available, the models should be regarded as extrapolations, and suffer in addition from some boundary effects due to procedures inherent to our methodology. Future use of the models should keep this in mind (in cases where needed, a larger modeled FOV can be constructed).\\


\section{Summary}
\label{sec:summ}

Observing 41 massive clusters with HST and Spitzer, the treasury RELICS program was primarily designed to find magnified high-z candidates, some of which are expected to be apparently bright enough for future spectroscopic follow-up from the ground and with JWST (e.g., \citealt{Salmon2017,Acebron2018}). SL models for the clusters are crucial for studying the magnified sources and, for example, for deriving their intrinsic brightness (Table \ref{tab:cand_z}), or star-formation properties, or for constructing the high-redshift luminosity function. 

In this work we have presented SL models for four RELICS clusters. Our analysis, based on the Light-Traces-Mass approach whose main advantage is its predictive power for finding multiple images, supplies the first published SL models for AS295 and for MACS0159. For A697 and MACS0025 we present improved models thanks to the new RELICS data. In our modeling we used as constraints both some previously known (in MACS0025, A697, AS295), and, importantly, several new identifications (in MACS0159, A697, AS295) of multiply imaged galaxies.

Out of the four clusters we analyzed, AS295 has the largest cumulative area of high magnification, possibly as large as that of the largest HFF clusters (Fig. \ref{areaMag}). This is much thanks to its two merging subclumps, which create a large region of high magnification between them (Fig. \ref{mags}; much of the region is outside each subclump's critical curves). Our analysis also indicates that MACS0025 is a highly magnifying lens, with  cumulative area of high magnification comparable to the typical HFF cluster (Fig. \ref{areaMag}). Also in the case of MACS0025 there is only a small critical area around each subclump, and most of high magnification area forms between the two subclumps. In that sense it should be noted that a large area of high magnification alone may not be sufficient for significantly enhancing the detection of high-redshift galaxies, and a sizable critical area  -- as was the case for the HFF clusters -- is also important. MACS0159 is the largest main-clump lens in the sample we analyze here, with an Einstein radius of $\theta_{\rm{E}} = 24.9 \pm 2.5"$ (for a source at $z_{s}=2$), and also has a significant, HFF-like area with high magnification. It is thus likely the most efficient lens amongst the four clusters we analyze here. A697, despite signatures of recent galaxy interaction with the BCG, or suggested merger scenarios along the line of sight (\citealt{macario2010,girardi2006}), seems fairly regular in terms of SL, and constitutes the smallest lens in our sample, as well as the one with smallest area of high magnification. 

We note that the above results, in particular the magnification power of the lenses, should be referred to as initial models which will be improved as additional model constraints become available.

To address the uncertainties in our SL models arising from the lack of spectroscopic information, we performed a suite of tests where we changed the initial configurations with respect to the redshift of the multiple images. We then examined the effect on the resulting density profiles and on the lens power (cumulative area of high magnification). 

In one set of trials we constructed new lens models with a range of input, main source redshifts that deviate from their fiducial photo-z value. In a second set of trials we allow systems with spectroscopic redshifts to vary, to check whether the redshifts are recovered, and probe the effect on the resulting models. Other tests were similarly designed to probe the effect of different redshift combinations in our models. For most of the cases the different density profiles agree well within the $3\sigma$ confidence level. Regarding the magnification values, our tests suggest that the redshift uncertainties typically propagate into differences of up to $50\%$ in the cumulative area of high magnifications (for $\mu > 5$ and $\mu > 10$). We also found that the spectroscopic redshifts in the couple of cases where available, were recovered to within 10\%.

We obtained GEMINI-N/GMOS spectroscopic observations to target multiply imaged galaxies and high-z candidates in the field of A697. While we did not identify any prominent feature in the multiple image spectra, we detect a single emission line in the spectrum of the high-z candidate Abell697-0636. Following its dropout nature and photometric redshift, we interpret this as a Ly$\alpha$ line, yielding a redshift of z$_{Ly\alpha} = 5.800 \pm 0.001$. The line is detected with $\sim15\sigma$ confidence, has an observed Gaussian width of $(117 \pm 15)$ km/s, and we measured a rest-frame equivalent width of ($52 \pm 22$) \AA.
 
The lens models we presented here, including mass-density, shear and magnification maps, are made available through the MAST archive.

\acknowledgments

We would like to thank the reviewer of this work for useful comments. This work is based on observations taken by the RELICS Treasury Program (GO-14096) with the NASA/ESA HST. Program GO-14096 is supported by NASA through a grant from the Space Telescope Science Institute, which is operated by the Association of Universities for Research in Astronomy, Inc., under NASA contract NAS5-26555. This work was performed in part under the auspices of the U.S. Department of Energy by Lawrence Livermore National Laboratory under Contract DE-AC52-07NA27344. RCL acknowledges support from an Australian Research Council Discovery Early Career Researcher Award (DE180101240). Based in part on observations obtained at the Gemini Observatory, which is operated by the Association of Universities for Research in Astronomy, Inc., under a cooperative agreement with the NSF on behalf of the Gemini partnership: the National Science Foundation (United States), the National Research Council (Canada), CONICYT (Chile), Ministerio de Ciencia, Tecnolog\'{i}a e Innovaci\'{o}n Productiva (Argentina), and Minist\'{e}rio da Ci\^{e}ncia, Tecnologia e Inova\c{c}\~{a}o (Brazil). Data were retrieved through the Gemini Observatory Archive and processed using the Gemini IRAF package. We thank Michael Lundquist for very useful help with the Gemini data reduction. 

\bibliographystyle{apj}
\bibliography{test}

\clearpage
\newpage
\mbox{~}
\clearpage
\newpage

\appendix

\section{Multiple image systems}
\label{appenTab}

\begin{table*}[h]
	\caption{Multiple images and candidates for MACS J0025.4-1222.}            
	\label{table:0025}      
	\centering   
    {\renewcommand{\arraystretch}{1.45}
	\begin{tabular}{c c c c c c c}        
		\hline\hline                 
		Arc ID & R.A. & Dec & $z_{\rm{spec}}$ & $z_{\rm{phot}}$\tablenotemark{a} & $z_{\rm{model}}$\tablenotemark{b} & individual RMS\tablenotemark{c}\\  
		&(J2000)&(J2000)& & & [$95\%$ C.I.]& (")  \\
		\hline          
		1.1 & 00:25:27.684 & -12:22:11.19 & 2.38\tablenotemark{d} & $2.57^{+0.16}_{-0.19}$ & - & 0.53\\ 
        1.2 & 00:25:27.162 & -12:22:21.69 & " & - & - & 0.62\\
        1.3 & 00:25:27.183 & -12:22:33.89 & " & - & - & 0.67\\
        1.4 & 00:25:27.524 & -12:22:23.50 & " & $2.36^{+0.21}_{-0.23}$ & - & 0.55\\
        \hline
        2.1 & 00:25:34.103 & -12:23:11.56 & - & $0.38^{+0.03}_{-0.04}$ & 3.80\tablenotemark{e} & 0.39\\
        2.2 & 00:25:34.096 & -12:23:14.37 & - & $3.82^{+0.13}_{-0.13}$ & " & 0.35\\
        2.3 & 00:25:34.068 & -12:23:15.38 & - & $3.78^{+0.15}_{-0.18}$ & " & 1.03\\
        \hline
        c3.1 & 00:25:31.900 &-12:23:06.28 & -& -& $\sim 1.9$ &- \\
        c3.2 & 00:25:31.900 &-12:23:05.01 & -& -&" &- \\
        c3.3 & 00:25:32.160 & -12:22:58.13 & - & $1.34^{+1.58}_{-1.12}$ &" &- \\
        \hline
        c4.1 & 00:25:29.274 & -12:22:42.57 & -& $2.72^{+0.28}_{-0.15}$ & \tablenotemark{f} &- \\
        c4.2 & 00:25:29.292 & -12:22:42.12 & -& -&" &- \\
        \hline
        c5.1 & 00:25:28.350 & -12:22:23.07 & -& $2.43^{+0.10}_{-0.24}$ & $\sim 2.2$ &- \\
        c5.2 & 00:25:28.334 & -12:22:22.26 & -& -&" &- \\
        c5.3 & 00:25:27.812 & -12:22:39.42 & -& -&" &- \\ 
		\hline\hline   
        \vspace{-1cm}
        \tablenotetext{1}
        {Photometric redshift from BPZ, taken from the RELICS catalog. Uncertainties correspond to $2\sigma$.}
         \tablenotetext{2}
        {Redshift prediction from the best-fit model. For candidate systems this value corresponds to the best prediction given by the model.}
        \tablenotetext{3}
        {Between observed and predicted location of the multiple images.}
        \tablenotetext{4}
        {Spectroscopic redshift reported in \citet{bradac08}}
        \tablenotetext{5}
        {Mean $z_{\rm{phot}}$ between images 2.2 and 2.3, fixed in the model.}
      \tablenotetext{6}
     {Poorly constrained.} 
	\end{tabular}}
    
\vspace{10mm}

	\caption{Multiple images and candidates for MACS J0159.8-0849.} 
	\label{table:0159}      
	\centering    
    {\renewcommand{\arraystretch}{1.45}
	\begin{tabular}{c c c c c c c}        
		\hline\hline                 
		Arc ID & R.A. & Dec & $z_{\rm{spec}}$ & $z_{\rm{phot}}$\tablenotemark{a} & $z_{\rm{model}}$\tablenotemark{b} & individual RMS\tablenotemark{c}\\  
		&(J2000)&(J2000)& & & [$95\%$ C.I.]& (")  \\
		\hline          
		 1.1 & 01:59:50.749 & -8:50:21.66 & - & $1.68^{+0.09}_{-0.27}$  & 1.55\tablenotemark{d} & 1.57\\
         1.2 & 01:59:48.306 & -8:49:57.57 & - & $1.41^{+0.30}_{-0.01}$ & " & 1.87\\
         1.3 & 01:59:49.267 & -8:49:58.49 & - & $0.49^{+0.05}_{-0.07}$ & " & 0.50\\
         \hline
         2.1 & 01:59:50.875 & -8:50:18.72 & - & $0.05^{+2.75}_{-0.03}$ & 1.55\tablenotemark{d} & 1.29 \\
         2.2 & 01:59:48.329 & -8:49:59.34 & - & $0.84^{+0.29}_{-0.08}$ & " & 0.65\\
         2.3 & 01:59:49.196 & -8:49:57.89 & - & - & " & 0.18\\
         \hline
         3.1 & 01:59:49.846 & -8:49:41.11 & - & $3.59^{+0.19}_{-0.07}$ & $2.63_{-0.01}^{+0.73}$ & 0.32 \\
         3.2 & 01:59:49.882 & -8:50:37.75 & - & $4.05^{+0.14}_{-0.24}$ & " & 0.66 \\
         \hline
         4.1 & 01:59:49.000 & -8:50:13.63 & - & $2.20^{+0.30}_{-0.07}$ & $1.46_{-0.11}^{+0.03}$ & 0.26 \\
         4.2 & 01:59:48.784 & -8:49:26.71 & - & $2.54^{+0.14}_{-0.33}$ & " & 0.80 \\
         \hline
         c5.1 & 01:59:50.911 & -8:49:59.05 & -&$0.91^{+1.84}_{-0.77}$ & $\sim 1.4$ &- \\
         c5.2 & 01:59:50.882 & -8:49:55.27 & -& $0.91^{+2.0}_{-0.78}$ &" &- \\
         \hline
         c6.1 & 01:59:49.543 & -8:50:08.64 & -& -& $\sim 2.3$ &- \\
         c6.2 & 01:59:49.127 & -8:49:13.75 & -& $2.36^{+0.16}_{-0.17}$&" &- \\
		\hline\hline   
        \tablenotetext{1}
        {Photometric redshift from BPZ, taken from the RELICS catalog. Uncertainties correspond to $2\sigma$.}
        \tablenotetext{2}
        {Redshift prediction from the best-fit model. For candidate systems this value corresponds to the best prediction given by the model.}
        \tablenotetext{3}
        {Between observed and predicted location of the multiple images.}
        \tablenotetext{4}
        {Mean $z_{\rm{phot}}$ between images 1.1 and 1.2 set to be the main source redshift, fixed in the model.}
	\end{tabular}}
 \end{table*} 
    
\begin{table*}[h]
	\caption{Multiple images and candidates for Abell S295.}       
	\label{table:S295}      
	\centering     
    {\renewcommand{\arraystretch}{1.45}
	\begin{tabular}{c c c c c c c}        
		\hline\hline                 
		Arc ID & R.A. & Dec & $z_{\rm{spec}}$ & $z_{\rm{phot}}$\tablenotemark{a} & $z_{\rm{model}}$\tablenotemark{b} & individual RMS\tablenotemark{c} \\  
		&(J2000)&(J2000)& & &[$95\%$ C.I.] & (")  \\
		\hline          
		1.1 & 02:45:24.123 & -53:01:51.05 &0.93\tablenotemark{d}& $0.84^{+0.03}_{-0.14}$ & - & 0.74 \\
        1.2 & 02:45:24.455 & -53:01:42.52 & " & - & -& 0.82\\
        1.3 & 02:45:25.128 & -53:01:37.57 & " & - & -& 0.34\\
        \hline
        2.1 & 02:45:24.102 & -53:01:50.01 & 0.93\tablenotemark{d} &-& - & 1.10\\
        2.2 & 02:45:24.361 & -53:01:42.84 & " & - & - & 0.91\\
        2.3 & 02:45:25.128 & -53:01:36.68 & " & - & - & 0.19\\
        \hline
        3.1 & 02:45:24.213 & -53:01:51.42 & 0.93\tablenotemark{d} & -&- & 0.72\\
        3.2 & 02:45:24.495 & -53:01:43.22 & " & - & - & 1.01\\
        3.3 & 02:45:25.234 & -53:01:37.68 & " & - & - & 0.43\\
        \hline
        4.1 & 02:45:24.093 & -53:01:47.34 & 0.93\tablenotemark{d}&- & - & 1.50\\
        4.2 & 02:45:24.181 & -53:01:44.38 & " & - & - & 0.72\\
        4.3 & 02:45:25.137 & -53:01:35.26 & " & $1.08^{+1.51}_{-0.40}$ & - & 0.47\\
        \hline
        \ 5.1\tablenotemark{e} & 02:45:34.874 & -53:03:05.50 & - & $0.84\tablenotemark{f}^{+0.02}_{-0.02}$ & $1.52_{-0.07}^{+0.13}$  &0.63\\
        5.2 & 02:45:35.142 & -53:03:04.81 & -& -& " &0.83\\
        5.3 & 02:45:37.287 & -53:02:47.51 &- & -& " &1.01\\
        \hline
       \ 6.1\tablenotemark{e} & 02:45:34.685 & -53:02:39.42 & - & $0.33^{+3.07}_{-0.23}$ & $2.17_{-0.49}^{+0.01}$  & 0.83\\
        6.2 & 02:45:34.466 & -53:02:40.80 &- & $3.25^{+0.25}_{-2.91}$ & " & 0.91\\
        6.3 & 02:45:33.468 & -53:02:48.00 & -& $2.74^{+0.38}_{-2.70}$ & " & 0.57\\
        \hline
        c7.1 & 02:45:34.812 & -53:02:41.94 & -& &$\sim 2.4$ &- \\
        c7.2 & 02:45:33.035 & -53:02:54.42 & -&$0.91^{+0.18}_{-0.38}$ &- &- \\
		\hline\hline   
        \vspace{-1cm}
        \tablenotetext{1}
        {Photometric redshift from BPZ, taken from the RELICS catalog. Uncertainties correspond to $2\sigma$.}
        \tablenotetext{2}
        {Redshift prediction from the best-fit model. For candidate systems this value corresponds to the best prediction given by the model.}
        \tablenotetext{3}
        {Between observed and predicted location of the multiple images.}
        \tablenotetext{4}
        {Spectroscopic redshift reported in  \citep{edge1994,Williams1999}}
        \tablenotetext{5}
        {While we refer to systems 5 \& 6 as secure, we acknowledge that due to lack of internal details their identification should be taken with somewhat more caution.}
        \tablenotetext{6}
        {This estimate is given by a poor SED fit.}
	\end{tabular}}     
      
\vspace{10mm}
	\caption{Multiple images for Abell 697.}   
	\label{table:697}      
	\centering     
    {\renewcommand{\arraystretch}{1.45}
	\begin{tabular}{c c c c c c c}        
		\hline\hline                 
		Arc ID & R.A. & Dec & $z_{\rm{spec}}$ & $z_{\rm{phot}}$\tablenotemark{a} & $z_{\rm{model}}$\tablenotemark{b} & individual RMS\tablenotemark{c}\\  
		&(J2000)&(J2000)& & & [$95\%$ C.I.]&  (")  \\
		\hline          
		1.1 & 08:42:57.098 & +36:22:03.38 & - & $1.54^{+0.30}_{-0.25}$ & 2.0\tablenotemark{d} & 0.73 \\  
        1.2 & 08:42:56.664 & +36:21:56.43 & - & $1.91^{+0.59}_{-0.75}$ & " &  0.25 \\  
        1.3 & 08:42:57.509 & +36:21:54.87 & -& - & " &  0.54 \\ 
        1.4 & 08:42:58.841 & +36:22:06.24 & - & $2.18^{+0.56}_{-0.33}$ & " & 1.35 \\
		\hline
        2.1 & 08:42:56.983 & +36:21:46.67 & - & $1.90^{+0.58}_{-0.57}$ & $1.96_{-0.28}^{+0.06}$ & 0.38 \\
        2.2 & 08:42:57.828 & +36:21:48.42 & - & - & " & 0.44\\
        2.3 & 08:42:58.789 & +36:21:55.99 & - & $2.40^{+0.19}_{-0.66}$ & " & 1.33 \\
        \hline    
        3.1 & 08:42:57.139 & +36:21:47.34 & - & $2.78_{-0.46}^{+0.37}$ & $2.97_{-0.38}^{+0.04}$ & 0.85 \\
        3.2 & 08:42:57.380 & +36:21:47.52 & - & $2.94_{-0.39}^{+0.42}$ & " & 1.15 \\
        p3.3\tablenotemark{e} & 08:42:58.905 & +36:21:59.237 & - & - & " & - \\
        \hline\hline 
        \tablenotetext{1}
        {Photometric redshift from BPZ (using a new run of \textsc{SExtractor} with parameters edited manually). Uncertainties correspond to $2\sigma$.}
        \tablenotetext{2}
        {Redshift prediction from the best-fit model.}
        \tablenotetext{3}
        {Between observed and predicted location of the multiple images.}
        \tablenotetext{4}
        {Mean $z_{\rm{phot}}$ between images 1.2 and 1.4 set to be the main source redshift, fixed in the model.}
        \tablenotetext{5}
        {Counter image predicted by the best-fit model but not identified in the observed image.}
	\end{tabular}} 
\end{table*}

\clearpage
\newpage

\section{Reproduction of multiple images}
\label{stamps}

\begin{figure*}[h]
\centering
\includegraphics[width=0.55\linewidth]{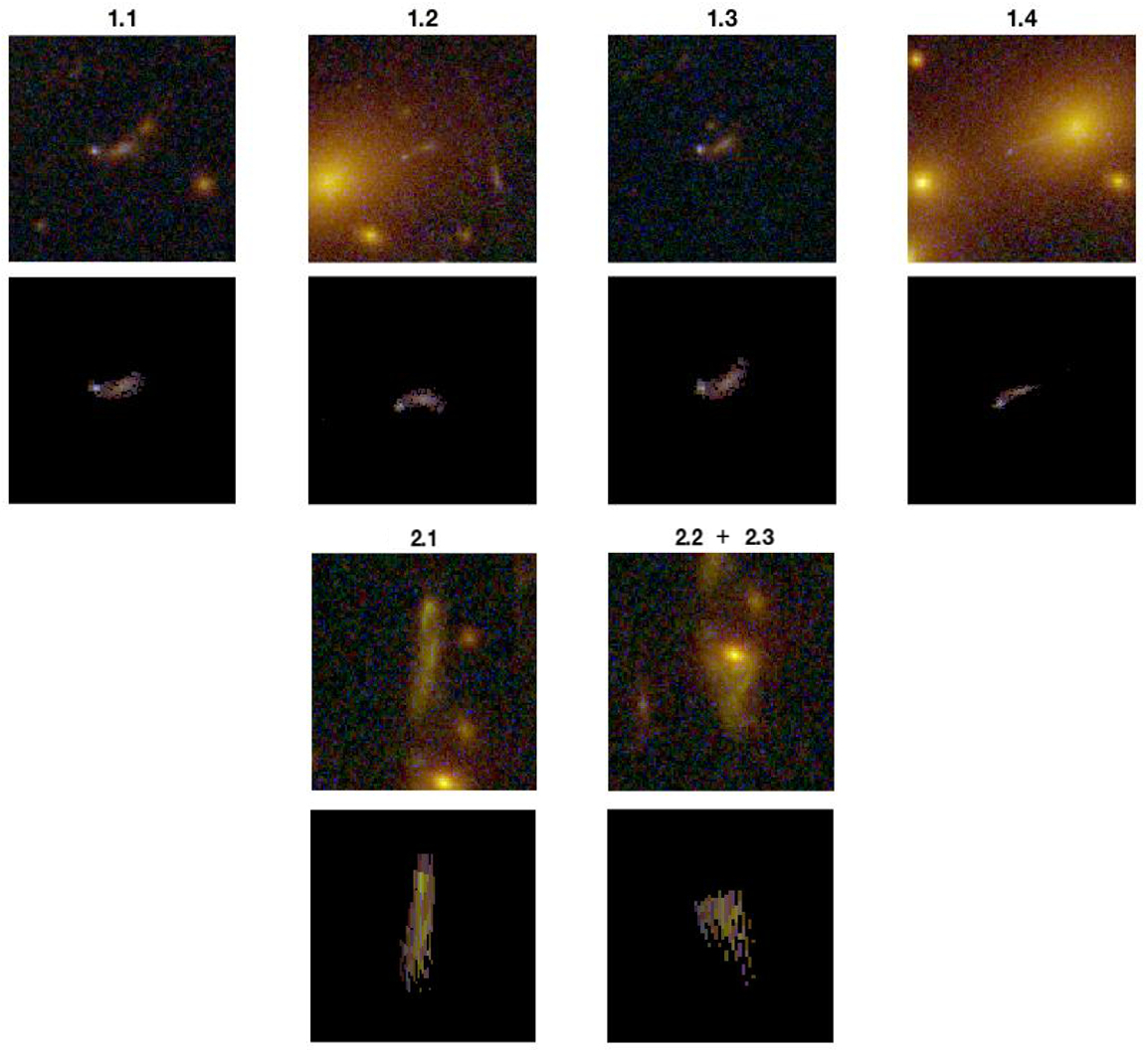} 
 	\caption{Multiple images reproduced by our best-fit model for MACS0025. For each system we de-lens the first counter image to the source plane and re-lens back to the image plane. The orientation, location and details of the predicted images (bottom rows) are similar to the observed images (upper rows), especially for system 1. The second and third counter images from system 2 are shown together. The prediction for system 2 follows the observed disposition of the counter images but with a somewhat different position/configuration. The images correspond to regions of approximately $5\arcsec \times 5\arcsec$, and small shifts may have been applied to center the reconstructed multiple-images on the observed positions.}
	\label{stamp_0025}
\end{figure*} 

\begin{figure*}[h]
\centering
\includegraphics[width=0.70\linewidth]{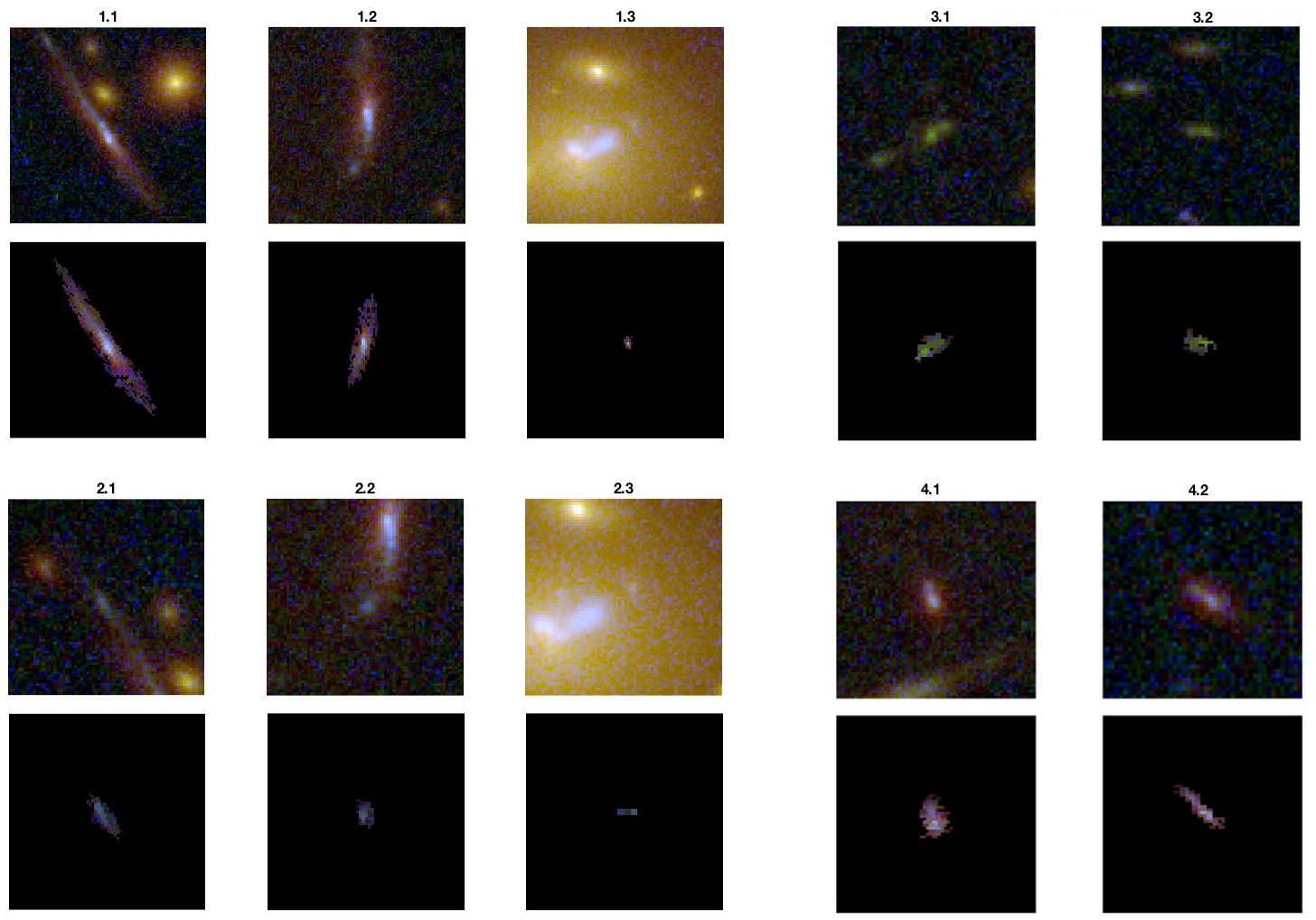} 
 	\caption{Multiple images reproduced by our best-fit model for MACS0159. For each system we de-lens the first counter image to the source plane and then re-lens back to the image plane. The orientation, location and details of the predicted images (bottom rows) are comparable to the observed images (upper rows) and validate our identification. The images correspond to regions of approximately $5\arcsec \times 5\arcsec$, and small shifts may have been applied to center the reconstructed multiple-images on the observed positions.}
	\label{stamp_0159}
\end{figure*} 

\begin{figure*}
\centering
\includegraphics[width=0.68\linewidth]{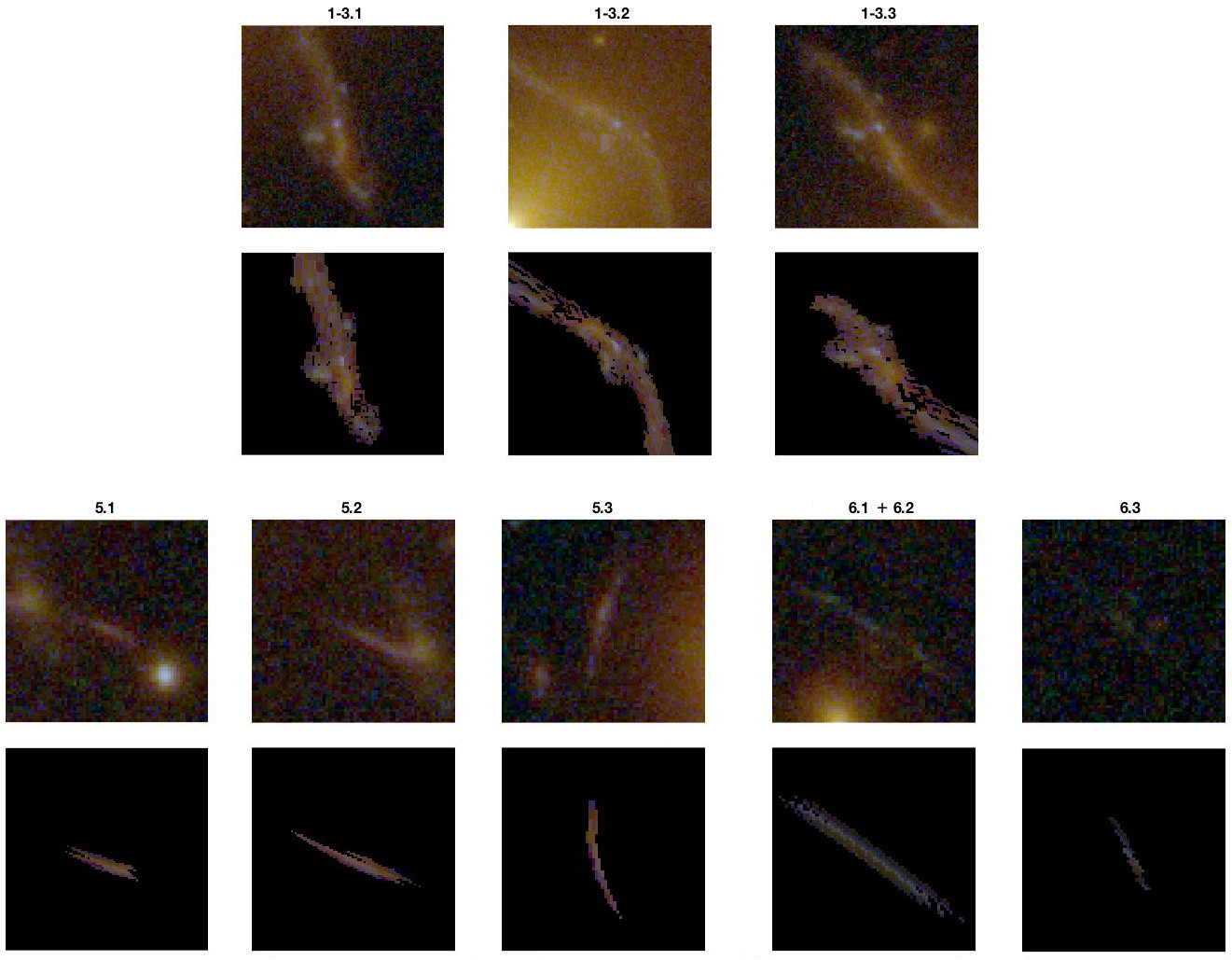} 
 	\caption{Multiple images reproduced by our best-fit model for AS295. We display systems 1 to 3 together and omit system 4 for sake of clarity, as they are all part of the giant arc. For each system we de-lens the first counter image to the source plane and re-lens back to the image plane. The orientation, location and details of the predicted images (bottom rows) are similar to the observed ones (upper rows). The images correspond to regions of approximately $5\arcsec \times 5\arcsec$, and small shifts may have been applied to center the reconstructed multiple-images on the observed positions.}
	\label{stamp_295}
\end{figure*} 

\begin{figure*}
\centering
\includegraphics[width=0.68\linewidth]{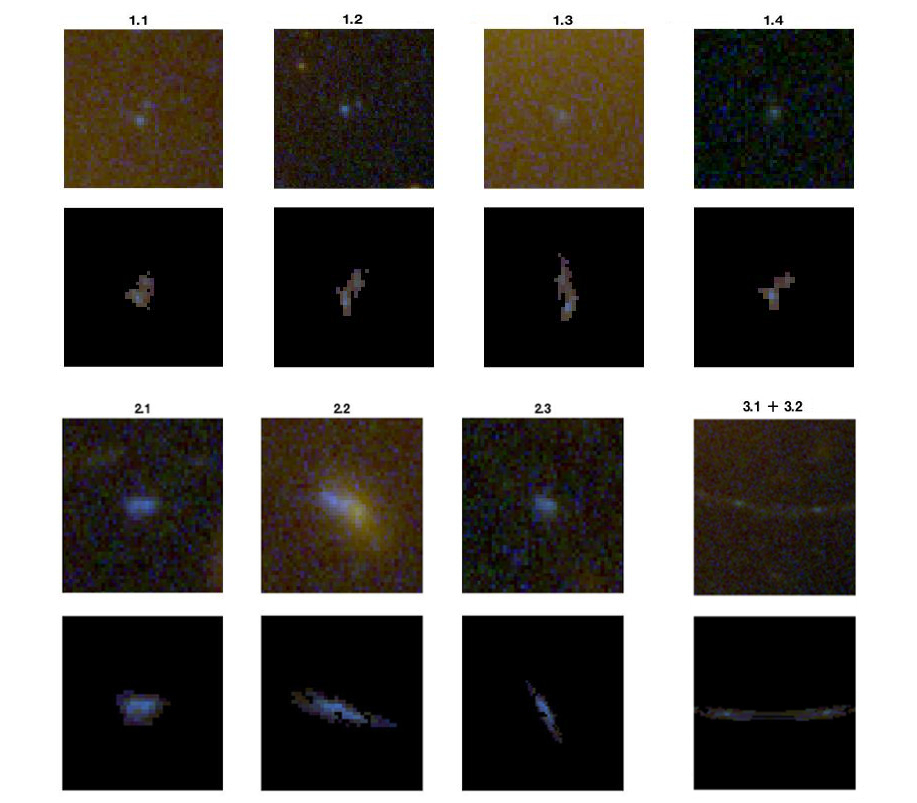} 
 	\caption{Multiple images reproduced by our best-fit model for A697. For each system we de-lens the first counter image to the source plane and then re-lens back to the image plane. The orientation, location and details of the predicted images (bottom rows) are comparable to the observed images (upper rows) and validate our identification. The images correspond to regions of approximately $5\arcsec \times 5\arcsec$, and small shifts may have been applied to center the reconstructed multiple-images on the observed positions.}
	\label{stamp_697}
\end{figure*} 

\clearpage
\newpage

\section{Photometric redshift uncertainties - effects on the mass-density profile}
\label{tests-z}

\begin{figure*}[h]
\centering
\includegraphics[width=.45\linewidth]{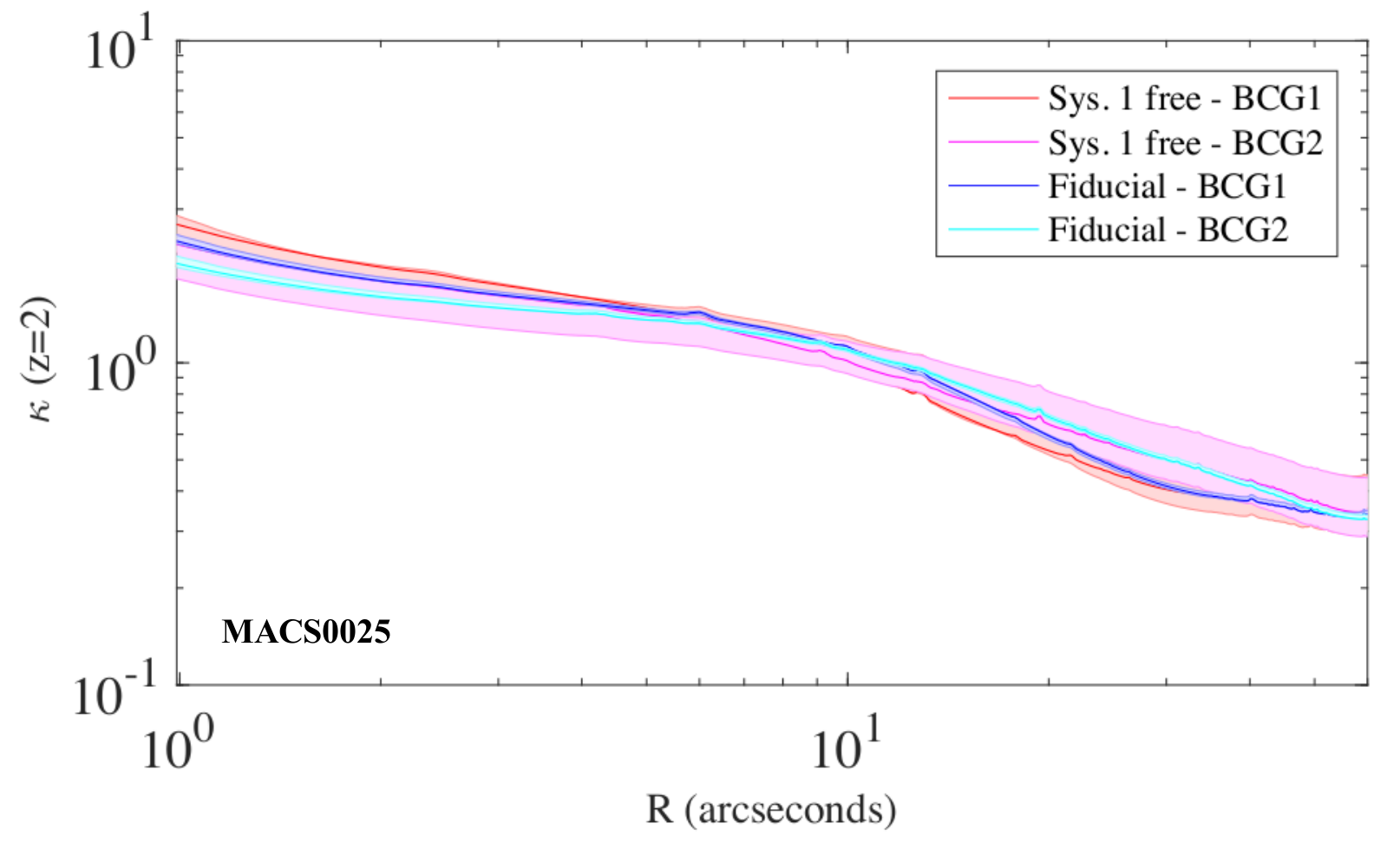}
	\includegraphics[width=0.45\linewidth]{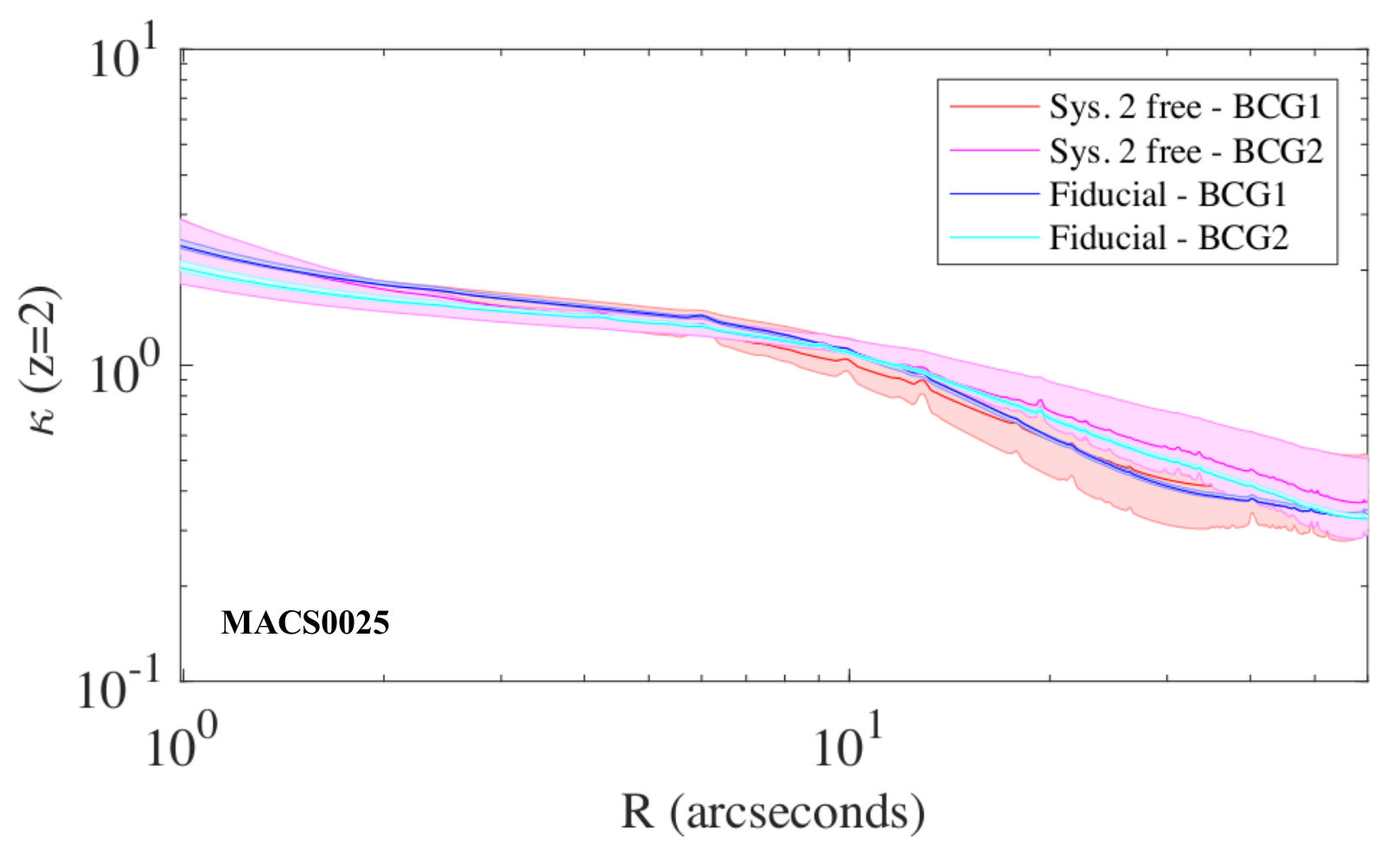} 
    \caption{Comparison of different radial mass-density profiles scaled to $z = 2$ for the cluster MACS0025. The fiducial model, where the redshifts of the two system of multiple images were fixed, is compared to the trial models. On the left panel the tested model has the redshift of system 1 left free, and one the right panel the redshift of system 2 is optimized in the model. Shaded regions correspond to the $68\%$ confidence level.}
   	\label{0025profs}
\end{figure*} 

\begin{figure*}[h]
\centering
    \includegraphics[width=0.45\linewidth]{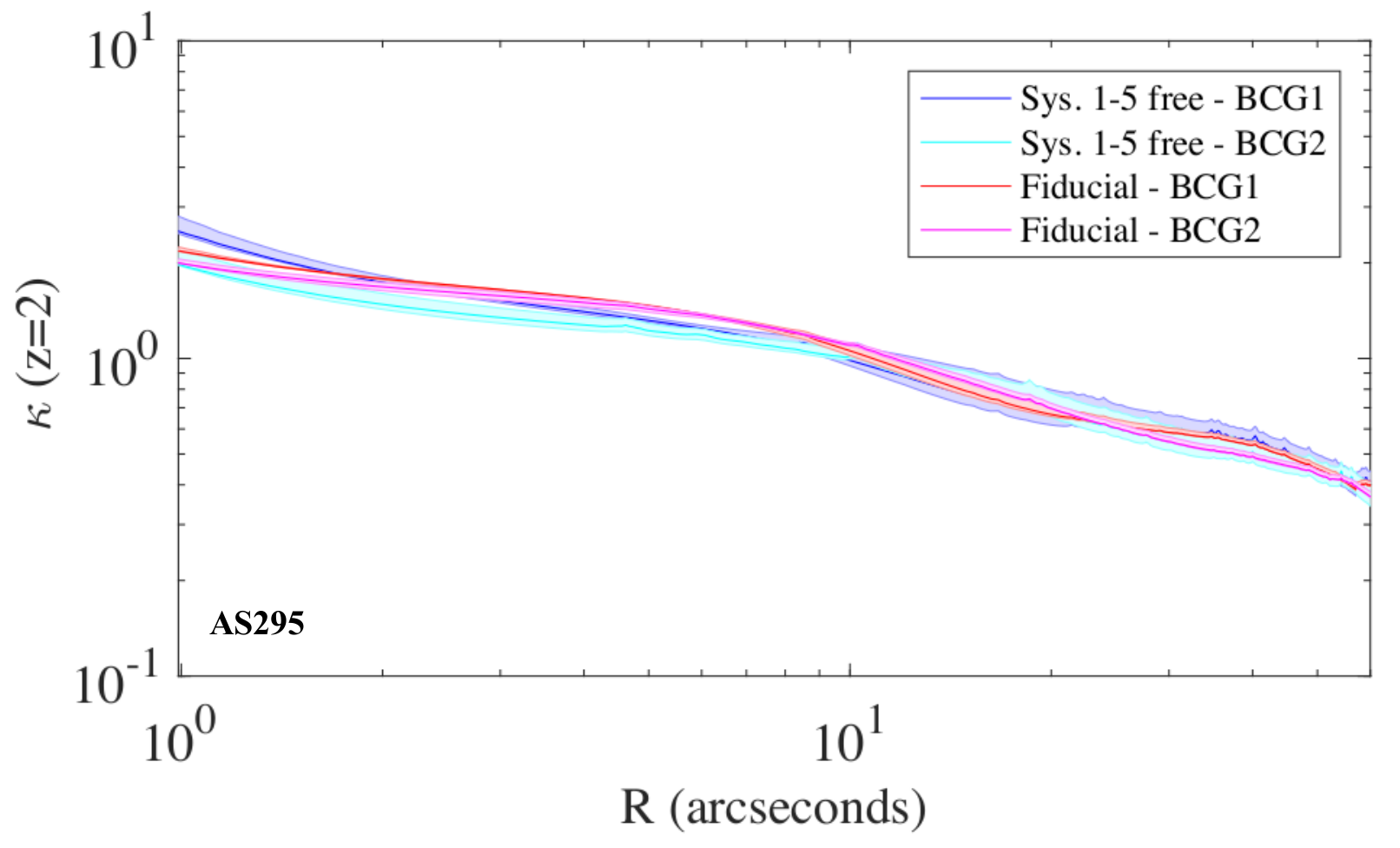}
    \caption{Comparison of different radial mass-density profiles scaled to $z = 2$ for the cluster AS295. The initial model has the redshift of systems 1-4 fixed to the spectroscopic value, while in the trial model these are free parameters. The redshift of system 5 is optimized in both models. Finally system 6 is left free in the fiducial model and fixed for the test. Shaded regions correspond to the $68\%$ confidence level.}
 	\label{295profs}
\end{figure*} 

\begin{figure*}[h]
\centering
\includegraphics[width=0.45\linewidth]{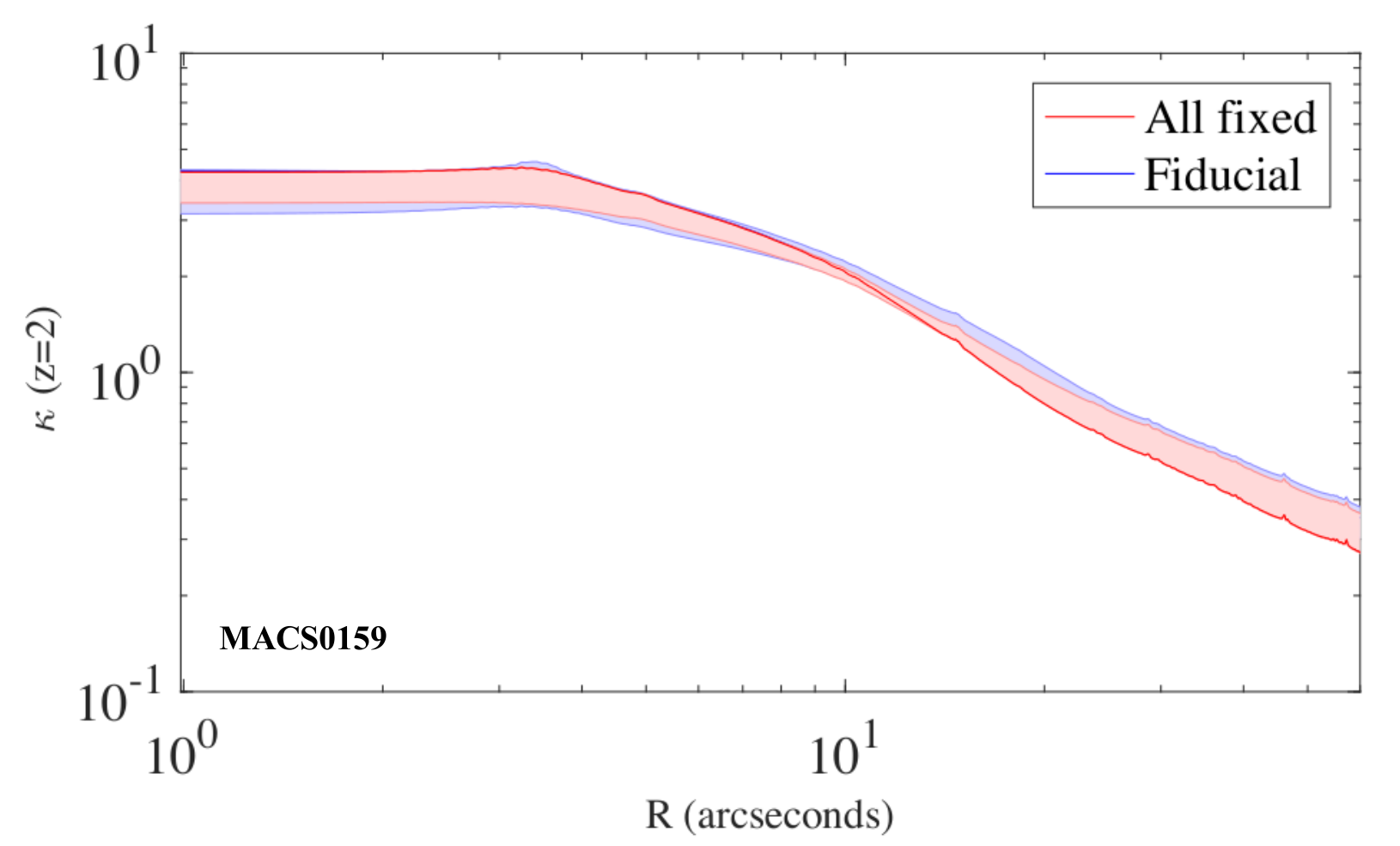} 
\includegraphics[width=0.45\linewidth]{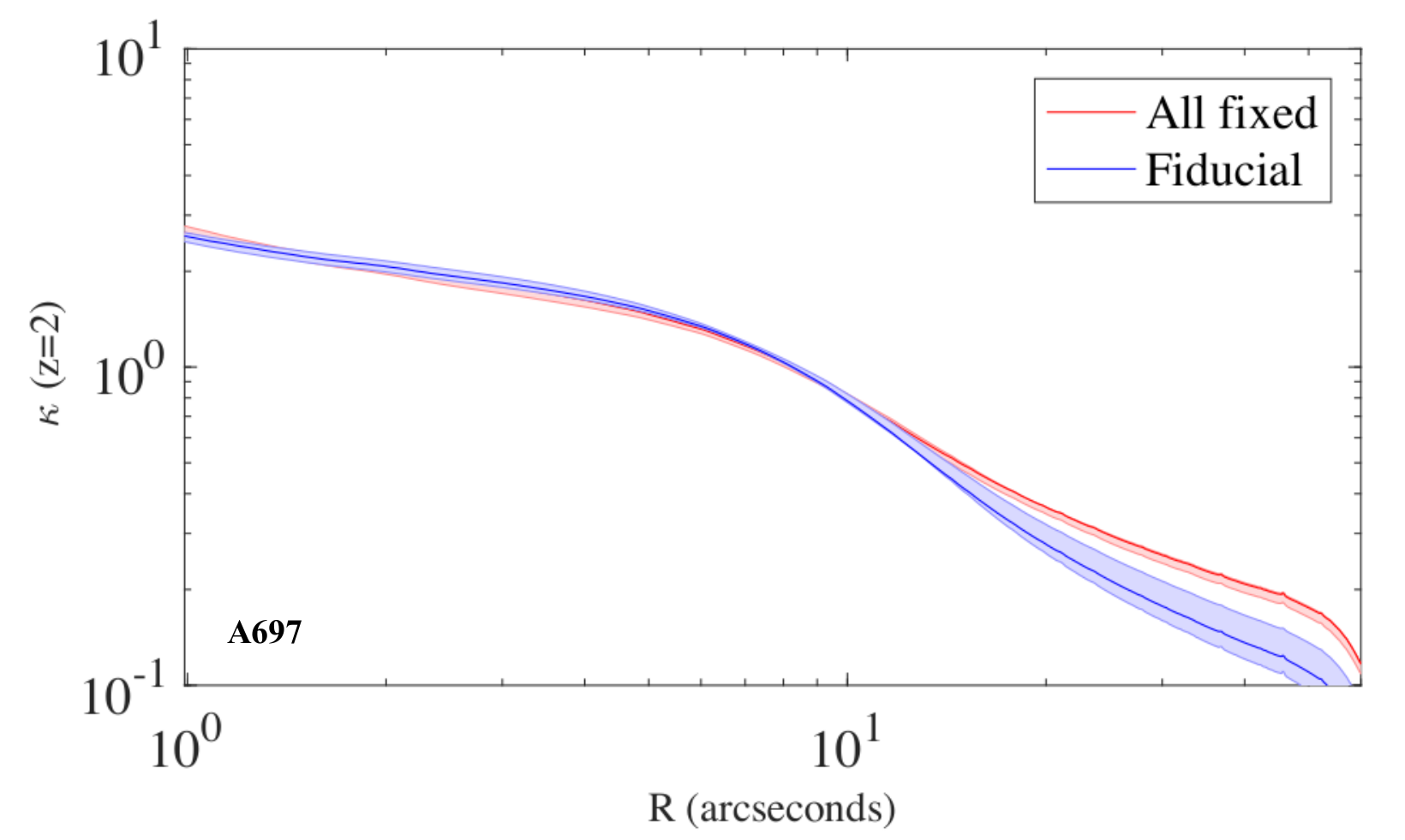}
\caption{Comparison of different radial mass-density profiles scaled to $z = 2$ for the clusters MACS0159 and A697. We show the original model, where most of the redshifts are free parameters, compared to the tested model where the redshift of all multiple images are set to the photometric estimates.  Shaded regions correspond to the $68\%$ confidence level.}
	\label{0159_697profs}
\end{figure*}

\end{}


\end{document}